\documentclass[12pt]{article}
\usepackage[utf8]{inputenc}
\usepackage{graphicx}
\graphicspath{{figures/}}
\usepackage{color}
\usepackage{amsmath}
\usepackage{float}
\usepackage{hyperref}
\usepackage{comment}
\usepackage[margin=1in]{geometry}
\usepackage[style=nature,doi=false,isbn=false,url=false]{biblatex}
\usepackage[labelfont=bf]{caption}

\usepackage[utf8]{inputenc}
\usepackage[T1]{fontenc}
\usepackage{mathptmx}
\usepackage{multirow}
\usepackage{chemformula}
\usepackage{upgreek}
\usepackage{float}
\usepackage{subcaption}
\usepackage{cleveref}
\usepackage[labelfont=bf, format=plain]{caption}
\include{MyCommand}

\title{{\bf Predicting Relative Populations of Protein Conformations without a Physics Engine Using AlphaFold2}}

\author{{\bf Gabriel Monteiro da Silva and Jennifer Y. Cui} \\
\textit{Brown University Department of Molecular Biology, Cell Biology, } \\
\textit{and Biochemistry, Providence, RI, USA} \\ \\
{\bf David C. Dalgarno} \\
\textit{Dalgarno Scientific LLC, Brookline, MA, USA} \\ \\
{\bf George P. Lisi and Brenda M. Rubenstein} \\
\textit{Brown University Department of Molecular Biology, Cell Biology,} \\
\textit{and Biochemistry} \\
\textit{Brown University Department of Chemistry} \\
\textit{Providence, RI, USA} \\ \\
}

\begin{document}
\maketitle



\section{Abstract}
This paper presents a novel approach for predicting the relative populations of protein conformations using AlphaFold 2, an AI-powered method that has revolutionized biology by enabling the accurate prediction of protein structures. While AlphaFold 2 has shown exceptional accuracy and speed, it is designed to predict proteins' single ground state conformations and is limited in its ability to predict fold switching and the effects of mutations on conformational landscapes. Here, we demonstrate how AlphaFold 2 can directly predict the relative populations of different conformations of proteins and even accurately predict changes in those populations induced by mutations by subsampling multiple sequence alignments. We tested our method against NMR experiments on two proteins with drastically different amounts of available sequence data, Abl1 kinase and the granulocyte-macrophage colony-stimulating factor, and predicted their relative state populations with accuracies in excess of 80\%. Our method offers a fast and cost-effective way to predict protein conformations and their relative populations at even single point mutation resolution, making it a useful tool for pharmacology, analyzing NMR data, and studying the effects of evolution.

\section{Introduction \label{intro}}

Proteins are essential biomolecules that carry out a wide range of functions in living organisms. Understanding their three-dimensional structures is critical for elucidating their functions and designing drugs that target them \cite{Roberston_ChemRev}. Historically, experimental techniques such as X-ray crystallography, nuclear magnetic resonance (NMR) spectroscopy, and electron microscopy have been used to determine protein structures \cite{Bai_2015,Su_2014,W_thrich_1990}. However, these methods can be time-consuming, technically challenging, and expensive, and may not work for all proteins \cite{Slabinski_2007}. To meet this challenge, \textit{ab initio} structure prediction methods, which use computational algorithms to predict protein structures from their amino acid sequences, have been developed \cite{Leach_2017}. For many years, \textit{ab initio} structure prediction methods have relied on physics-based algorithms that leverage statistical mechanics to calculate the most stable protein structures \cite{O_dziej_2005}. Although highly successful, these methods are limited by their steep computational expense, and consequently, have struggled to predict the structures of larger and more complex proteins \cite{Jothi_2012}.

The recent development of machine learning algorithms has significantly improved the speed of protein structure prediction \cite{Torrisi_2020, AlQuraishi_2021}. One of the most remarkable achievements in this area is the AlphaFold 2 (AF2) engine developed by DeepMind, which uses a deep neural network to predict ground state protein structures from amino acid sequences \cite{jumper_highly_2021,tunyasuvunakool_highly_2021}. AlphaFold 2 was trained using large amounts of experimental data and incorporates co-evolutionary information from massive metagenomic databases \cite{jumper_highly_2021}. Its accuracy has revolutionized the field of protein structure prediction \cite{Baek_2023,jumper_highly_2021,Lin_2023}, opening up new possibilities for drug discovery and basic research with clear consequences for human health \cite{Roney_2022, Callaway_2022}. 

However, studies have found that the default AF2 algorithm is limited in its capacity to predict alternative protein conformations and the effects of sequence variants \cite{Chakravarty_2022}. Although AF2's inability to predict multiple conformations is unsurprising given its initial scope, the capacity to make predictions of different conformations would be as revolutionary as the capacity to accurately predict ground states. Phenomena that involve different conformations such as fold-switching and order-disorder transitions are ubiquitous across the proteome \cite{Porter_2018,Bryan_2010} and are directly tied to the activity of many macromolecules  \cite{Kim_2021}. 
Moreover, methods that can rapidly predict multiple conformations may have the potential to revolutionize drug discovery by uncovering substantially more drug targets \cite{Borkakoti_2023}. To fully realize this potential, methods like AF2 will need to account for the relative populations of different conformations (states) since the conformational equilibrium of drug receptors is directly related to their affinities for drugs \cite{Xie_2020, Michielssens_2015}. A prime example of this relationship is Imatinib \cite{Iqbal_2014}, a tyrosine inhibitor whose exceptional selectivity for Abl1 kinase was found to be caused by the enzyme's significant preference for conformational states that facilitate Imatinib binding and subsequent induced-fit steps \cite{Wilson_2015}.  

In an attempt to realize this potential, researchers have devised new ways of employing the AF method to detect conformational changes, with significant success in a few test cases \cite{Wayment_Steele_2022,Vani_2023,Meller_2023,Stein_2022,del_alamo_sampling_2022}. Although AF2 conventionally fails to predict fold-switching, researchers have found that sub-sampling the input multiple sequence alignments (MSAs) and increasing the number of predictions leads to structural ensembles that capture different physiologically-relevant conformations from the same sequence \cite{Wayment_Steele_2022}. These predictions can be used as seeds in molecular dynamics (MD) simulations seeking to explore larger swaths of the conformational space and the relative populations of each predicted state \cite{Vani_2023}. Despite being a significant improvement over methods that only predict ground states, methods such as these still rely on expensive MD simulations to infer most relative state population information, which comes at a significant cost compared to simply running a prediction engine. 

Here, we show that these MD simulation steps may be unnecessary if the goal is to discover major alternative conformations and their relative populations in a high-throughput fashion, such as for contrasting differences in the dynamics of orthologs of a protein of interest. We take inspiration for this work from the observation that proteins from the same evolutionary line can have differences in relative state populations that are strongly correlated with the genetic distance between them \cite{Wilson_2015}. Since AF2 works by decoding co-evolutionary signals \cite{Roney_2022} and previous research has suggested that subsampling MSAs leads to accurate predictions of different conformations of the same protein \cite{Wayment_Steele_2022}, it seems reasonable to hypothesize that the instructions for fold-switching between major states could be decoded from sequence data alone. If this hypothesis holds true, AF2 and other AI-based methods could be capable of quantifying sequence-encoded fold-switching signals, which would make it possible to predict not only alternative conformations of the same protein but also changes in its relative state populations stemming from point mutations. 

\begin{figure}[H]
\centering
    \includegraphics[width=525pt]{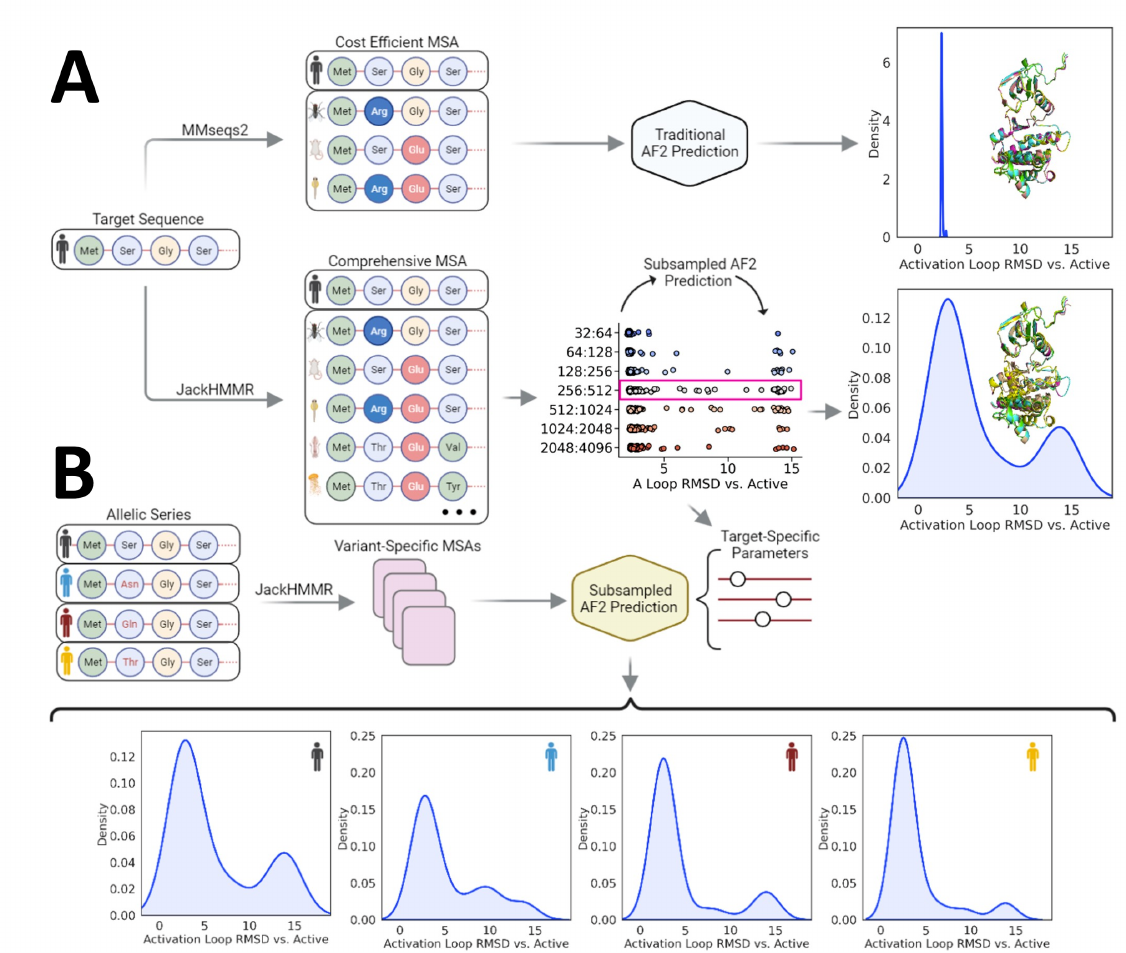}
    \caption{Summary of the subsampled AF2 workflow employed in this study. (A) Traditionally, AF2 predicts the structure of a target sequence by using a multiple sequence alignment (MSA) that is generated by methods that generally seek to balance sequence count, diversity, and speed. When running AF2 with such MSAs, the predicted structures are often similar to each other even with a large number of independent predictions (seeds). (B) In this study, we show that subsampling comprehensive MSAs causes AF2 to output predictions that occupy different conformations of the same protein, and the predicted frequency of each conformation based upon a range of random seeds strongly correlates with its experimentally-determined relative state population. In particular, we illustrate that subsampled AF2 is capable of predicting relative state populations of evolutionarily-related kinase cores and dynamics of the GMCSF protein with single substitution resolution.}
     \label{fig_1}
\end{figure}

With this as motivation, we show how subsampling multiple sequence alignments can generate ensembles of protein conformations (see Figure \ref{fig_1} for an overview of our method) and systematically test AF2's capacity to predict differences in the populations of conformers in two illustrative proteins: the Abl1 tyrosine kinase core and the granulocyte-macrophage colony-stimulating factor (GMCSF). Diverging from previous works, as a first example, we focused on detecting changes in the active state population across the Src kinase to Abl1 evolutionary line and tested our method's ability to predict the effects of single and double point mutations known or suspected to shift state distributions. Crucially, we found that subsampled AF2 can qualitatively predict both the positive and negative effects of mutations on the active state populations of kinase cores with up to eighty percent accuracy. We also found that AF2 predicts most of the activation loop intermediate states in the active-to-inactive transition of the kinase core with an ensemble that is comparable to that obtained from multi-microsecond MD simulations. As a second example, we predicted changes in the relative state populations of GMCSF, a protein with minimal known homology, in response to point mutations. Our predictions strongly correlated with experimentally-determined NMR chemical shifts, further showcasing subsampled AF2's remarkable capacity to decode signals pertaining to conformational dynamics even when sequence data is scarce. Altogether, these results highlight the strong, yet untapped potential of AF2 for predicting conformational ensembles of mutated and evolutionarily-related proteins, which will have substantial impacts on the fields of biophysics, drug design, and NMR.  

\section{Results \label{results}}

\subsection{Optimizing MSA Subsampling to Predict Kinase Core Conformational Ensembles}

\begin{figure}[H]
\centering
    \includegraphics[width=400pt]{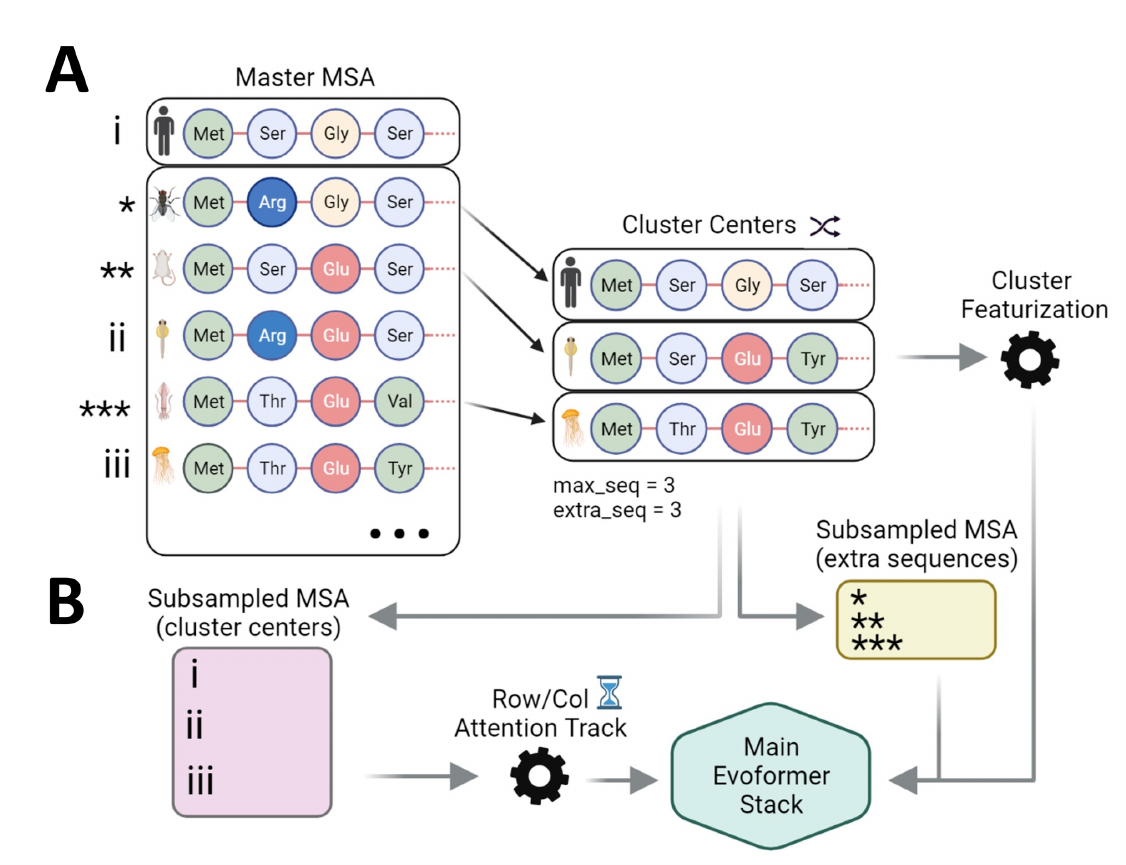}
    \caption{AF2's multiple sequence alignment (MSA) clustering heuristic. (A) An MSA of arbitrary length is built from a target sequence and passed to AF2, which randomly selects a number of sequences (defined by \emph{max\_seq}) from the input MSA. Each of the selected sequences becomes a cluster center around which the sequences not selected in the previous step are distributed. The target sequence is always selected as a cluster center. The clusters obtained through this process are featurized and relevant statistics are calculated. (B) All of the previously discussed elements are used by AF2 for inference. Cluster features and a number of random non-cluster-center sequences (defined by \emph{extra\_seq}) are processed and passed to the Main Evoformer Stack, while the MSA containing the cluster centers is processed, passed to the comparatively expensive row/col attention track, and then finally passed to the Main Evoformer Stack as well.}
     \label{fig_2}
\end{figure}

In recent years, multiple groups have observed that AF2 with different parameters and inputs from MSAs is capable of predicting conformational changes based on sequence data alone \cite{Vani_2023, Wayment_Steele_2022}. These alternative AF2 pipelines share the principle of subsampling MSAs to amplify co-evolutionary signals at different structural domains \cite{Wayment_Steele_2022}. In its standard implementation, AF2 takes as input a target sequence and a corresponding multiple sequence alignment. An arbitrary number of sequences (defined by the \textit{max\_seq} parameter) are randomly selected from the master MSA (the target sequence is always selected), and the remaining sequences are clustered around each of the selected sequences using a Hamming distance. Both the cluster centers and a sample from each cluster with a length of \emph{extra\_seq} are used by AF2 for inference (see Figure \ref{fig_2}). Previous works have shown that a significant reduction in the values of \emph{max\_seq} and \emph{extra\_seq} from their default values achieves ensemble prediction for a series of model systems \cite{Meller_2023}.

\begin{figure}[H]
\centering
    \includegraphics[width=350pt]{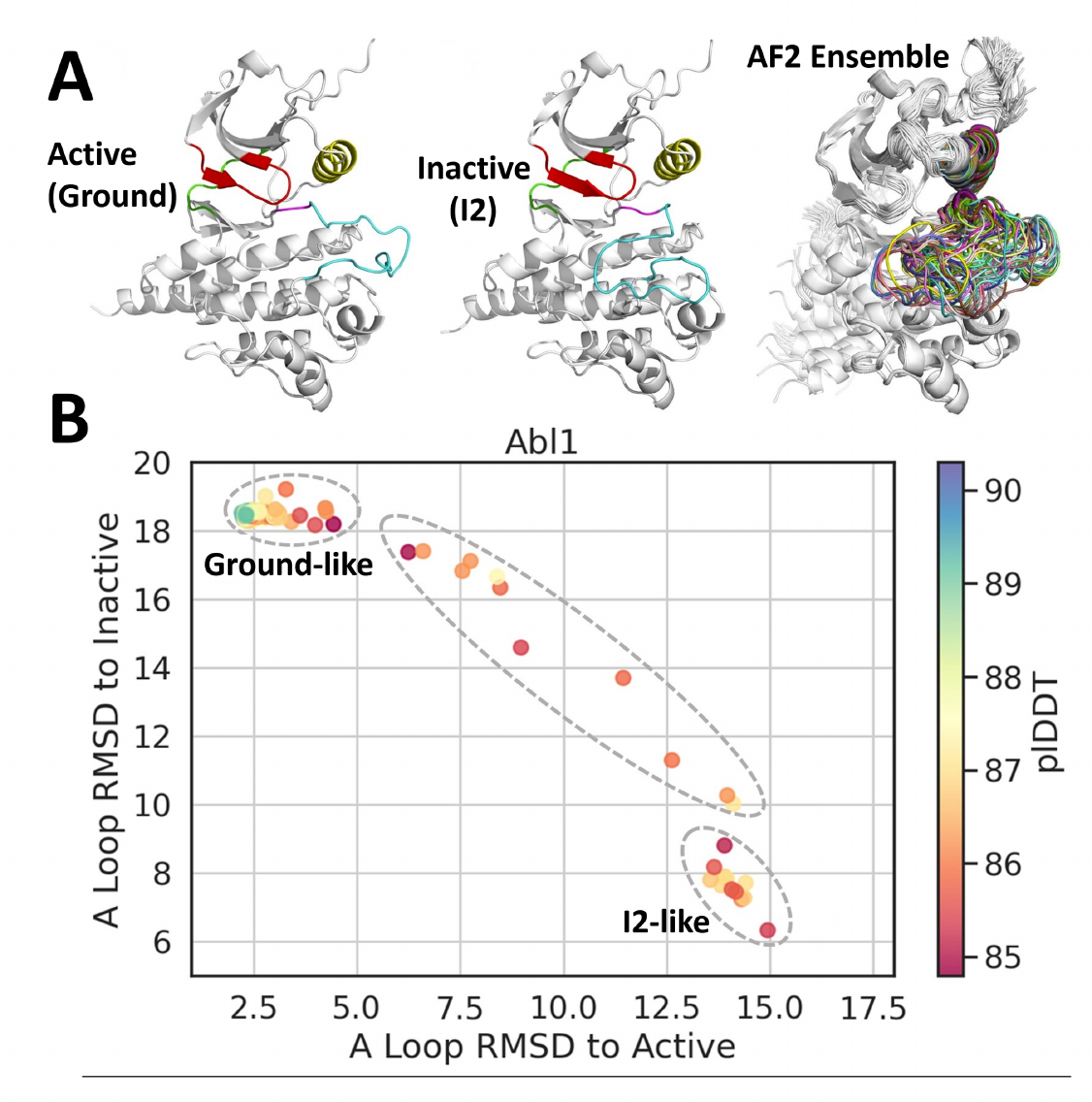}
    \caption{(A) Top: Three-dimensional models of the Abl1 kinase core in its active and inactive  conformations. Structural elements that are relevant for kinase function are represented in different colors. Cyan: activation loop; red: phosphate-binding loop; yellow: activation helix; green: hinge; and pink: DFG motif. Models are taken from the PDB (6XR6 for active core, 6RXG for inactive core) \cite{Xie_2020}. Top Right: Ensemble of 160 three-dimensional models of the Abl1 kinase core as predicted by subsampled AF2, with the activation loop highlighted in different colors for each prediction. (B) Two-dimensional projection of 160 Abl1 kinase core predictions from subsampled AF2 plotted according to their backbone RMSDs relative to the Abl1 kinase core active (6XR6) and inactive states (6XRG). Data points are colored according to their average predicted local distance difference test score (pLDDT) as calculated by AF2, which is a metric of the confidence/accuracy of AF2 predictions \cite{jumper_highly_2021}. Bottom Left: Cluster of predictions considered to be in the ground state. Bottom Right: Cluster of predictions considered to be in the inactive, I2 state.}
     \label{fig_3}
\end{figure}

Motivated by these observations, we started our work by systematically testing the accuracy of different AF2 parameter combinations for predicting the Abl1 kinase core structural ensemble. We chose the Abl1 kinase core (residues 235-497) as our first test case due to this system's extensively documented conformational dynamics \cite{Xie_2020}. Abl1 is thought to occupy three major conformations with different populations (see Figure \ref{fig_3}). In solution, Abl1 primarily exists in an active (ground) state, which as the name suggests, is capable of productive phosphorylation. Infrequently, Abl1 will switch to inactive state 1 (I1), and then to inactive state 2 (I2) \cite{Xie_2020}, which strongly binds to Imatinib (Gleevec) \cite{Wilson_2015}. While the change from ground to I1 is subtle and mostly involves shifting side-chain dihedrals, the transition from I1 to I2 involves considerable backbone rearrangements: the activation loop, composed of about 20 residues, detaches from its resting position below the C-helix and folds on itself, a change that involves the activation loop shifting by over 15 \r{A} from its original position as shown in Figure \ref{fig_3}.

The accuracy of our ensemble predictions was defined as their capacity to replicate the wild-type Abl1 kinase core conformations and their correct relative populations as validated by nuclear magnetic resonance experiments. Specifically, we sought a combination of parameters that led to an ensemble of predictions that met the following criteria: the ground state is the most frequent prediction within the ensemble, the transition from the ground to I2 state is captured within the ensemble, and the I2 state is present in the ensemble more frequently than transition states. Importantly, we opted to examine the relative populations of the ground and I2 states because of the large backbone rearrangement involved in the transition between these conformations, which is more likely to be reproduced by AF2 than the comparatively small dihedral flips in the ground to I1 transition. We optimized the accuracy achieved as a function of the following parameters: \emph{max\_seq}, \emph{extra\_seq}, number of seeds, and number of recycles (see Table S1 for a complete list of tests and parameters). We evaluated the ensemble resulting from each parameter set by measuring the activation loop backbone RMSD relative to either the active kinase core (PDB 6XR6) or the I2 kinase core (PDB 6XRG) for each prediction. This decision is rooted in the fact that the activation loop is the structural element that changes the most (in terms of backbone motions) upon the transition from the ground to I2 state \cite{Xie_2020}. 

To encourage AF2 to generate a full ensemble of Abl1 conformations, we started by compiling an extensive MSA spanning over 600,000 sequences using the JackHMMR algorithm \cite{Finn_2011} on wild-type Abl1 kinase core (residues 235-497) sequences pulled from the Uniprot90 \cite{Suzek_2014}, Small BFD \cite{Steinegger_2018}, and MGfny \cite{Richardson_2022} databases. 
To increase the statistical power of our results, we then ran 32 predictions with independent seeds for each test, and enabled dropouts during inference to sample from the uncertainty of the models. All other parameters were left in their default settings (4 recycles per prediction, 5 models per seed, a total of 160 predictions per run, 3 independent runs with unique seeds, 480 predictions per test). 

\begin{figure}[H]
\centering
    \includegraphics[width=350pt]{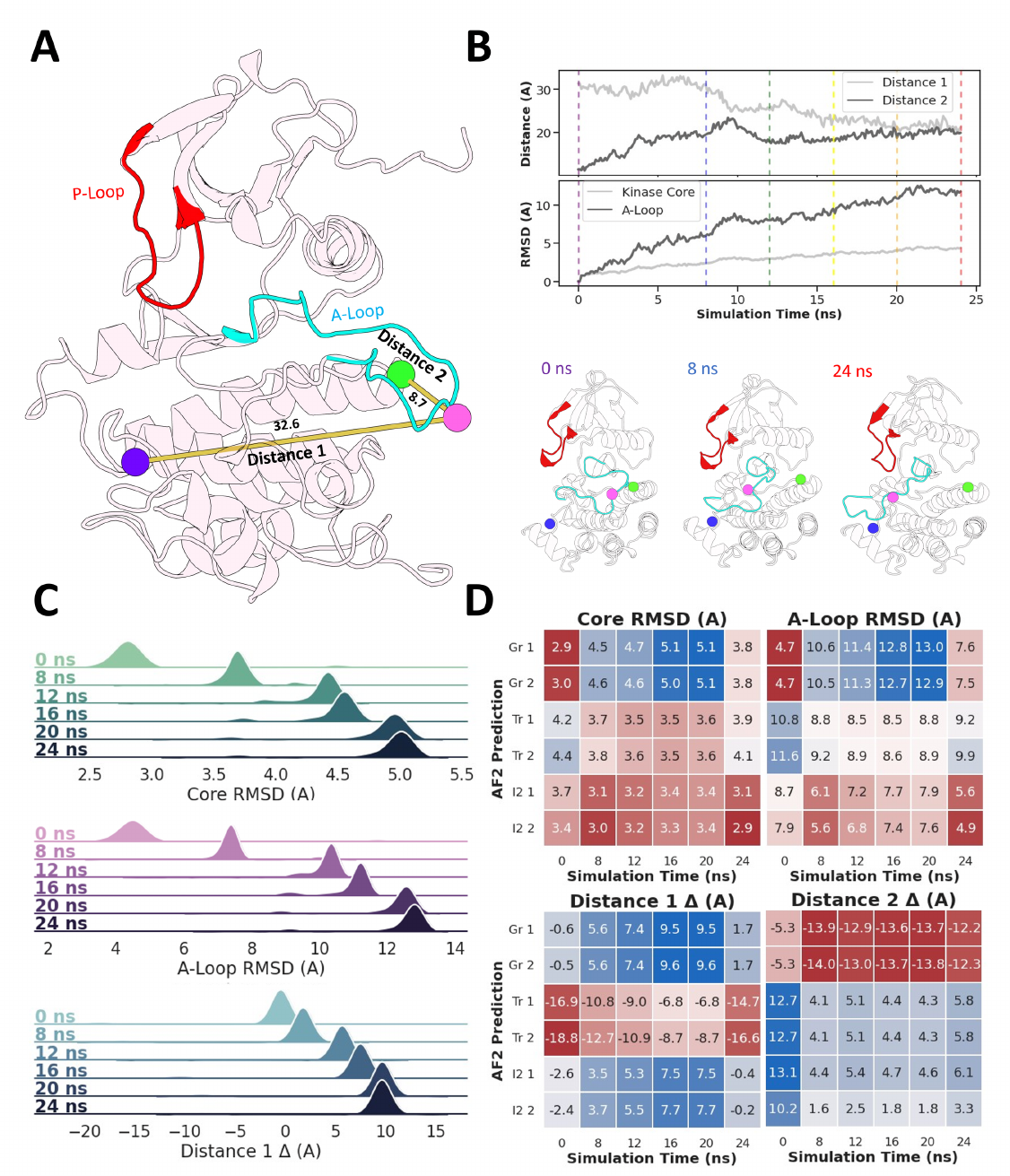}
    \caption{Comparison between the ground to I2 trajectory obtained using enhanced-sampling MD simulations of the Abl1 kinase core and representative AF2 predictions. (A) Structural elements relevant for Abl1 function and that are expected to shift significantly over the ground to I2 transition.  (B) Evolution of four relevant structural observables across the trajectory containing the ground to I2 transition. Vertical lines indicate representative snapshots extracted from the trajectory for downstream comparisons with AF2 predictions (top). Three representative snapshots from the MD trajectory at 0 (ground), 8 (transition), and 24 (I2) ns, respectively (bottom). (C) Distribution of three observables relative to the MD snapshots for 160 subsampled AF2 predictions. Core and A-Loop RMSDs are defined as the backbone RMSDs of each AF2 prediction's kinase core (residues 242 to 459) or activation loop (residues 379 to 395) vs. the kinase core or activation loop backbone of the MD snapshot selected at each time point. Distance deltas are defined as the difference in atom pair distances between each AF2 prediction and each respective MD snapshot. Distance 1 corresponds to the distance between the backbone oxygens of E377 and L409, and Distance 2 corresponds to the distance between the backbone oxygens of L409 and G457. (D) Comparison between six representative AF2 predictions and the six MD snapshots described above.}
     \label{fig_4}
\end{figure}

Consistent with previous observations, we found that merely changing \emph{max\_seq} and \emph{extra\_seq} was sufficient to obtain predictions that met all of the above criteria, with a \emph{max\_seq}:\emph{extra\_seq} ratio of 256:512 leading to the most diverse results in terms of activation loop conformations (see Figure \ref{fig_3}). Importantly, the ensemble of activation loop conformations predicted by AF2 with the above parameter set is distributed across the ground state to I2 state transition in Abl1, with no predictions falling outside the boundaries of known activation loop conformations and no blatantly unphysical or misfolded predictions. As a further test to the claim that we are actually predicting conformations along a transition, we compared the ensemble of 160 subsampled AF2 Abl1 predictions to representative snapshots extracted from a Ground to I2 trajectory generated with enhanced-sampling MD simulations of apo Abl1 in solution. Specifics about the methodology used to generate this trajectory are described in Appendix 1 of the Supporting Information. Representative results from this comparison are illustrated in  Figure \ref{fig_4} and the results of the entire analysis are illustrated in Figures S2-S5. Although we expected to observe a range of conformations, the coverage of the activation loop transition is remarkable and suggests the possibility of using AF2 to sample intermediate states and uncover pathways and mechanisms.

\begin{figure}[H]
\centering
    \includegraphics[width=425pt]{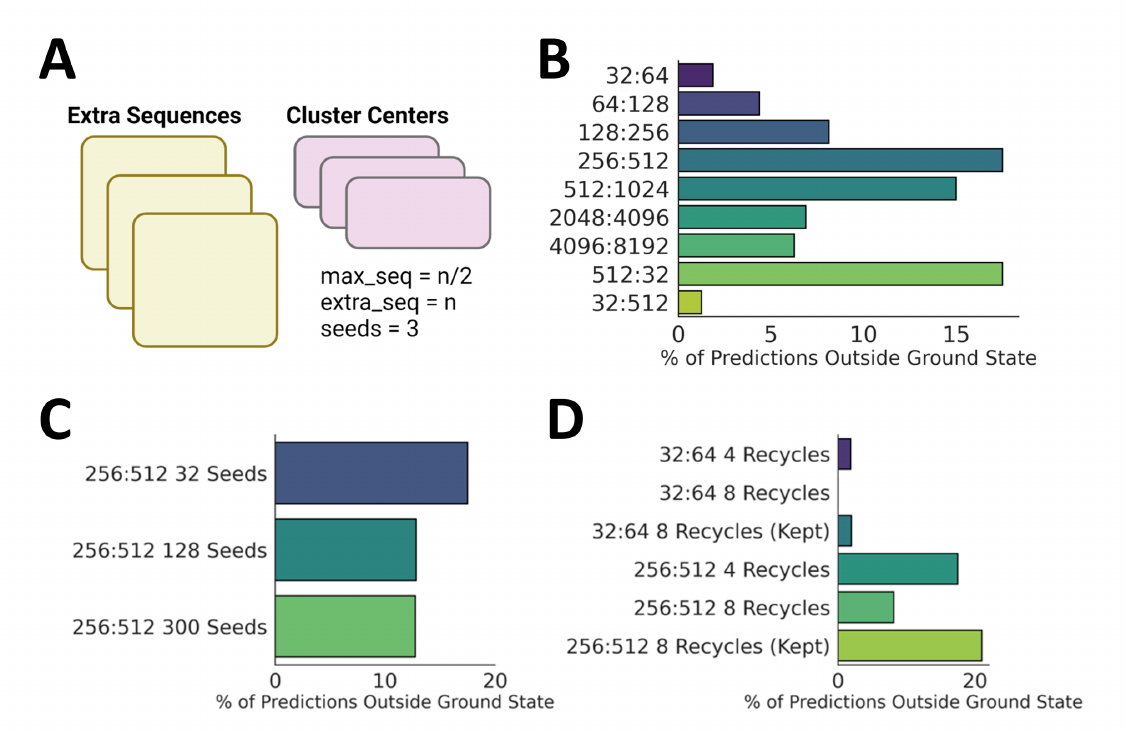}
    \caption{Percent of Abl1 kinase domain conformations predicted to fall outside of the ground state using AF2 based on different MSA clustering parameters (number of sequences selected as cluster centers, number of sequences sampled from the clusters, number of seeds used in the prediction, and number of recycles). (A) Summary of the MSA subsampling and clustering algorithm implemented in AF2. (B) Percent of Abl1 kinase core conformations predicted to fall outside of the ground state using subsampled AF2 based upon the number of seeds used and the amount of recycling performed, and whether recycling intermediates are kept. (C) Impact of changing the number of seeds (each seed corresponds to an independent AF2 prediction); and (D) impact of changing the number of recycles and if structures from recycled iterations are included in the analysis or discarded.}
     \label{fig_5}
\end{figure}

In order to better quantify the effects of each parameter change, we binned each predicted structure into three classes based on the backbone RMSD of relevant structural elements (activation loop, phosphate-binding loop, and C helix) with respect to the backbone of these elements in the ground (active) state, as defined by the lowest-energy structure assignment in the NMR ensemble PDB 6XR6 \cite{Xie_2020}. Since the RMSD with respect to the ground state of the majority of predictions clustered within 3 \r{A}, we classified predictions with RMSD values greater than 3.5 \r{A} as ``not in the ground state." Through this binning, we observed that the 256:512 and 512:1024 values for\textit{ max\_seq} and \textit{extra\_seq} led to predictions in which the ground state is populated 80\% and 85\% of the time, respectively. Of note, NMR results suggest that the relative state population of the Abl1 kinase core's ground state in solution is 88\%, which is in surprisingly good agreement with our AF2 predictions \cite{Xie_2020}. In contrast with the effects observed by changing MSA composition and length, increasing the number of seeds beyond 128 did not lead to significant changes in the state distribution, suggesting a degree of determinism in the prediction results, presumably stemming from AF2 training and information encoded in the co-evolutionary signal (see Figure \ref{fig_5}).

Interestingly, predictions with the \emph{max\_seq} and \emph{extra\_seq} parameters of 512 and 8, respectively, led to results that are similar to those of the 512:1024 test. Similarly, changing \emph{max\_seq} and \emph{extra\_seq} to 8:1024 led to results that closely resemble those from the 8:16 test. These results suggest that the \emph{max\_seq} parameter is the principal driver of alternative state predictions. This is unsurprising considering the different roles played by each parameter: the MSA of length defined by the \emph{max\_seq} argument and formed by the sequences randomly selected as cluster centers is passed to the expensive row/column attention Evoformer track, while the MSA of length \emph{extra\_seq} skips it. Due to the increased computational effort needed for featurization and attention, we expect AF2 to distill significantly more coevolutionary signal from the MSA of length \emph{max\_seq}, thus changes to \emph{max\_seq} will exert greater influence than changes to \emph{extra\_seq}.
        
Finally, we also tested the hypothesis that changing the number of recycles (\emph{n\_recycles}) per seed could lead to changes in predicted state distributions by doubling the number of recycles. Interestingly, increasing the number of recycles significantly increases the population of the ground state, suggesting that the recycling stage plays a role in AF2's propensity to generate different conformations. 
Considering all of the above, we defined our target-specific parameters for all subsequent kinase predictions as follows: \emph{max\_seq}: 256, \emph{extra\_seq}: 512, \emph{n\_recycles}: 4, \emph{n\_models}: 5, \emph{n\_seeds}: 96. Considering its significant impact on the distribution of predictions, the optimization of the \emph{max\_seq} parameter is paramount for successfully obtaining conformational ensembles when running AF2. While 256 cluster centers (defined by \emph{max\_seq} = 256) works for Abl1, significantly smaller values are likely to be required for protein systems with less available sequence data.

\subsection{Predicting the Conformations of Key Members of a Kinase Evolutionary Line Using Subsampled AF2 \label{family}}

\begin{figure}[H]
\centering
    \includegraphics[width=350pt]{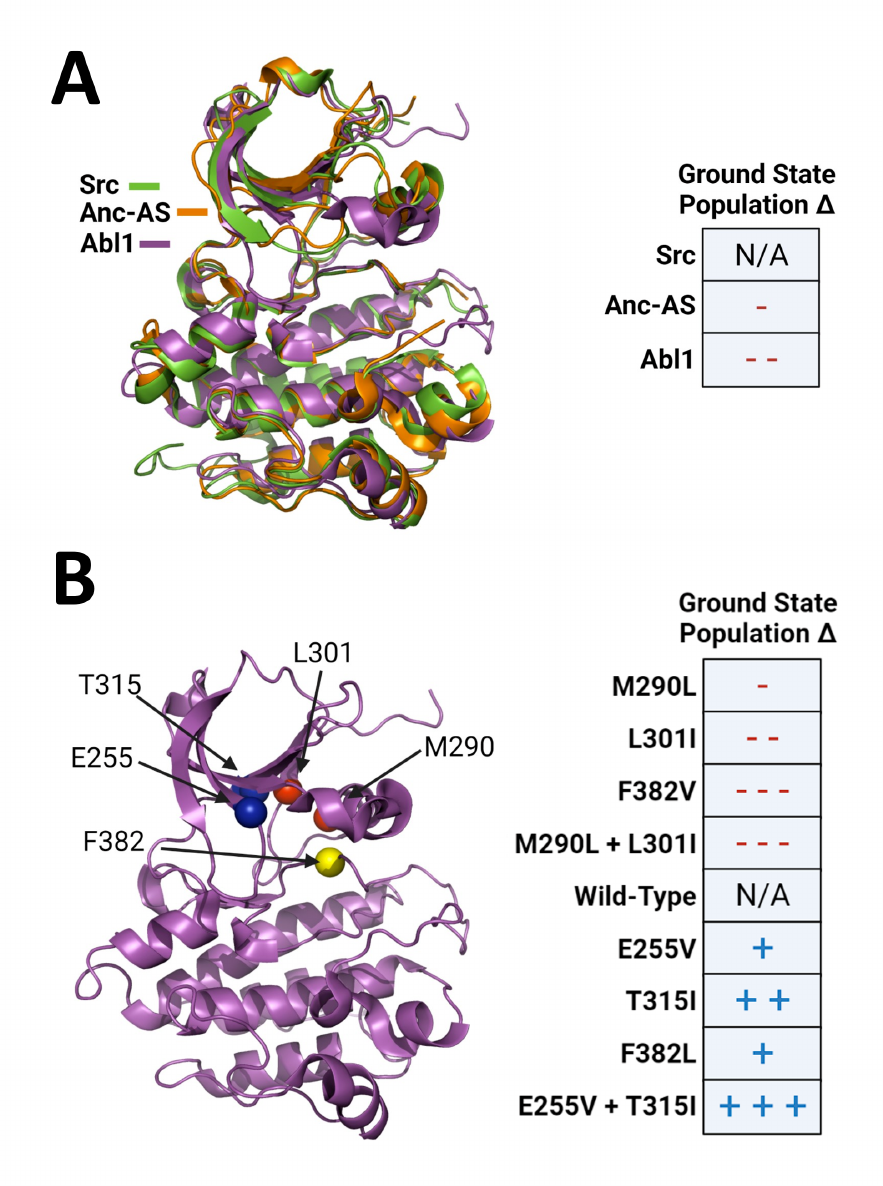}
    \caption{Summary of experimental observations regarding the relative state populations of kinase cores along Abl1's evolutionary line and in the Abl1 allelic series studied here. (A) Abl1, Anc-AS, and Src are part of the same evolutionary line and share significant sequence and structural homology, but Abl1 occupies the ground state significantly less frequently than Src, and Anc-AS occupies it slightly less than Src, suggesting that Abl1 evolution has directed it to be more flexible than Src \cite{Wilson_2015}. (B) A series of point mutations in wild-type Abl1 are known to increase or decrease the relative population of the enzyme's active (ground) state \cite{Xie_2020}. Residues with mutations known to increase the population of the ground state are shown as blue spheres in the Abl1 structure on the left side, while those with mutations known to decrease the ground state population are shown as red spheres. The yellow sphere denotes phenylalanine 382, part of the DFG motif, which can be mutated to valine to increase the ground state population, or to leucine to reduce it. While the effects of the E255V + T315I mutation on the ground state population were not directly reported in the literature, we used its cumulative effect on kinase activity and Imatinib binding \cite{Hoemberger_2020} to infer an increase in the ground state population.}
     \label{fig_6}
\end{figure}

Given our success predicting the relative populations of the Abl1 kinase core ground and I2 states, we next studied AF2's potential for predicting state distributions without the need for downstream MD simulations. As a basic sanity check, we tested if the Abl1 prediction results were actually the product of AF2 decoding coevolutionary signal pertaining to relative state populations, or just a fortuitous coincidence. Accordingly, we used the same subsampled AF2 protocol outlined in the previous section to predict the relative state populations of the wild-type Src kinase core, which is known to occupy the ground (active) state significantly more frequently than Abl1 \cite{Wilson_2015} (see Figure \ref{fig_6}), making it an attractive control case for this test. If our hypothesis regarding the potential of subsampled AF2 is indeed correct, we expect that the method will output significantly more predictions of ground state Src than it did for ground state Abl1, mirroring the experimentally-confirmed relative state distribution differences between the two kinases. Accordingly, we built a large MSA for the Src kinase core (residues 235-497) sequence using the same procedure as described for Abl1, and ran our implementation of subsampled AF2 with it as an input. We then measured the relative population of the Src kinase core ground and I2 states.

\begin{figure}[H]
\centering
    \includegraphics[width=400pt]{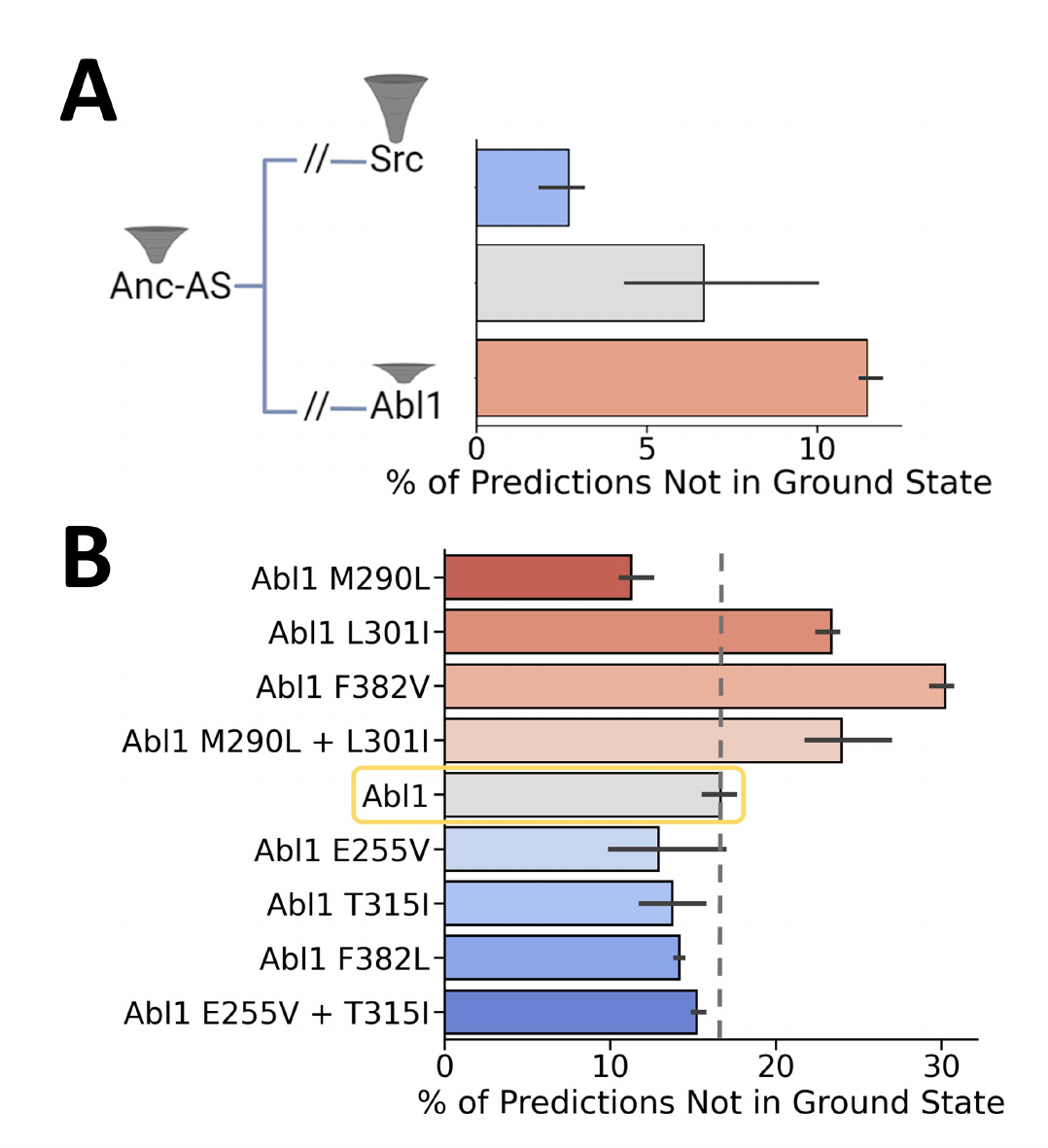}
    \caption{Subsampled AF2 predictions of the percent of conformations not in the ground state for proteins along the Src to Abl1 evolutionary pathway and Abl1 resistance-causing mutations. (A) Representation of the evolutionary history connecting Abl1, Anc-As, and Src, and the relative population of the active/ground state for each kinase core \cite{Wilson_2015}. Right: Percentage of predictions that fall outside of the ground state for each test case. (B) Results of subsampled AF2 for the Abl1 kinase domain allelic series. Data are colored based on the expected relative state populations of each mutation: red-tinted bars represent mutations that are known to decrease the ground state population, while blue-tinted bars represent mutations with the opposite effect \cite{Xie_2020}. Error bars were calculated from three independent sets of predictions, each with 32 unique seeds and 5 models, totaling 160 predictions per replicate.}
     \label{fig_7}
\end{figure}

Crucially, the vast majority of Src kinase core predictions from subsampled AF2 were found to be in the ground state, with a predicted relative state population of 97\% compared to 89\% for Abl1, as summarized in Figure \ref{fig_7}. Interestingly, none of the Src predictions were found to be in the I2 state, although the enzyme is known to infrequently occupy this conformation. This suggests a resolution limitation in using AF2 to predict relative state populations: conformations with very low occupancy such as I2 in Src might be missed by the algorithm in its current implementation. A potential cause of this limitation is the fact that AF2's prediction models were extensively trained on structures from the Protein Data Bank (PDB), in which most of the structures of Src and its orthologs are in the ground state \cite{Modi2021}.

Despite resolution caveats, subsampled AF2 correctly predicted the stark contrast in conformational distributions between the Abl1 and Src kinase cores, lending credence to its promise as a high-throughput method for predicting relative state populations. To further probe this capacity, we applied subsampled AF2 to make predictions for the Anc-AS kinase core (residues 1-263), an Abl1 ancestor with a known conformational state distribution and dynamics \cite{Wilson_2015}, and compared the results to the Abl1 and Src cases. The sequences of Abl1, Anc-AS, and Src used to generate the MSAs and used as target sequences for AF2 are summarized in Figure S1. While there are other Abl1 ancestors that could be used in this test, there are experimental results for Anc-AS, including a deposited structure in the Protein Databank (4UEU), justifying its choice. For the subsampled AF2 predictions to be considered accurate, the relative population of the ground state in the Anc-AS predictions should be in-between the populations of the same state for the Abl1 and Src predictions, as observed in experimental results. Once again, subsampled AF2 correctly replicated experiments (Figure \ref{fig_7}), as Anc-AS was predicted to be in the ground state 93\% of the time, in-between the frequencies predicted for Src (97\%) and Abl1 (89\%).

\subsection{Predicting the Conformations of a Kinase Allelic Series with AF2 \label{allelic}}

While the results we obtained for the Abl1 to Src evolutionary line are promising, an even more impactful application would be using it to predict how such populations change across an allelic series with single substitution resolution. This is because many point mutations in proteins are thought to lead to different phenotypes - such as drug resistance - by changing conformational landscapes and relative state populations. 

To measure the capacity of subsampled AF2 to fill this niche, we repeated our predictions using a series of Abl1 single and double mutants with well-characterized and significant effects on the relative populations of the ground and I2 states, and contrasted the results with those obtained from the wild-type prediction. Specifically, we tested the method on four mutations that are expected to decrease the population of the ground state (M290L, L301I, F382V, M290L + L301I), and four mutations that are expected to increase it (E255V, T315I, F382L, E255V + T315I). The mutations tested, their locations in Abl1, and their expected effects on ground state populations are summarized in Figure \ref{fig_6} (see Table S2 for the full list of mutations tested and their expected outcomes in the relative state populations of the Abl1 kinase core).  Importantly, the length of the kinase core sequence used as input for this test (229-515) differs slightly from the previous (235-497) so as to match the length of the constructs tested in the literature.  This difference caused a slight variation in the wild-type Abl1 ground state predictions (84\% vs. 89\%).

AF2 predictions for this allelic series are summarized in Figure \ref{fig_7}. Strikingly, subsampled AF2 correctly predicted a change in the relative state population and its direction in over 80\% of the tested cases. Although promising, these results are not without significant caveats. First: the M290L predictions are inaccurate. Specifically, the effects of the mutation on the ground state population are predicted to be the opposite of those seen in experiments. Second: the prediction accuracy only applies in a qualitative sense, as double mutations that are known to significantly increase the ground state population such as M290L + L301I are predicted by subsampled AF2 to increase it only slightly more than single mutations such as M301I, which is known to cause a more subtle increase. We believe that this inaccuracy is a direct consequence of the incorrect M290L prediction, which should also result in the underestimation of the effects of double mutations including M290L. Furthermore, the statistical significance of the results is reduced for the benchmarks where the mutations are known to reduce ground-state populations. We hypothesize that AF2 performed better at predicting decreases in the ground state population because its predictions reflect an abundance of Abl1 I2 structures in its training set.

\subsection{Predicting GMCSF Conformations with AF2 \label{GMCSF}}

\begin{figure}[H]
\centering
    \includegraphics[width=450pt]{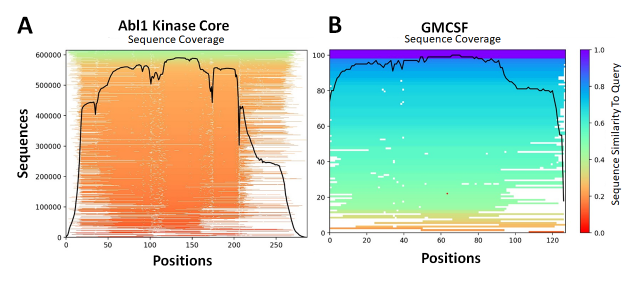}
    \caption{Comparison between MSA length and per-position coverage for (A) Abl1 and (B) GMCSF multiple sequence alignments used as input in AF2.}
     \label{fig_8}
\end{figure}

Considering the success of our Abl1 predictions, we sought to test if the accuracy of these predictions was contingent on the wealth of kinase sequence data, or if we could obtain similar results with significantly less sequence data. To do so, we repeated our prediction workflow but using the sequence of the human granulocyte-macrophage colony-stimulating factor (GMCSF). GMCSF is a 14 kDa monomeric glycoprotein that plays a central role in innate immunity, stimulating a variety of cells in response to pathogens \cite{Lee_2020}. In contrast to that for kinases, literature regarding GMCSF's structure and dynamics is sparse, and sequence data for homologs or orthologs is orders of magnitude less abundant. Crucially, the MSA built for the wild-type human GMCSF sequence is only 112 sequences long, while that for human Abl1 kinase is over 600,000 sequences using the same parameters and methods (see Figure \ref{fig_8}). This stark contrast represents a useful opportunity to study how the accuracy of subsampled AF2 is modulated by the availability of sequence data.

Previous NMR results have shown that the dynamics of GMCSF's N-terminal helix A are involved in the binding of heparin and other charged molecules \cite{Cui_2020}. Specifically, it is thought that the relative position of helix A with respect to the center of mass of the protein is flexible, usually being tightly packed against helix C through $\pi$-$\pi$  interactions among histidines 15, 83, and 87. Experimental results suggest that this closed configuration is the most stable GMCSF conformation when packed as a crystal \cite{Walter_1992}. Thus, for clarity, this closed conformation will be henceforth referred to as the ground state. It is thought that the $\pi$-$\pi$  interactions are eventually weakened either through intrinsic breathing motions and resultant interactions with the solvent or via induction by other molecules, and helix A is thought to move away from the GMCSF core. This opening motion should expose part of the GMCSF core to solvent and creates a V-shaped groove that is thought to be the binding site of heparin and other immune system modulators \cite{Cui_2020}. This binding-competent GMCSF state will be referred to as the open conformation. Importantly, GMCSF binds heparin much more strongly in an acidic environment, suggesting that helix A dynamics are protonation-dependent \cite{Cui_2020}.

\begin{figure}[H]
\centering
    \includegraphics[width=400pt]{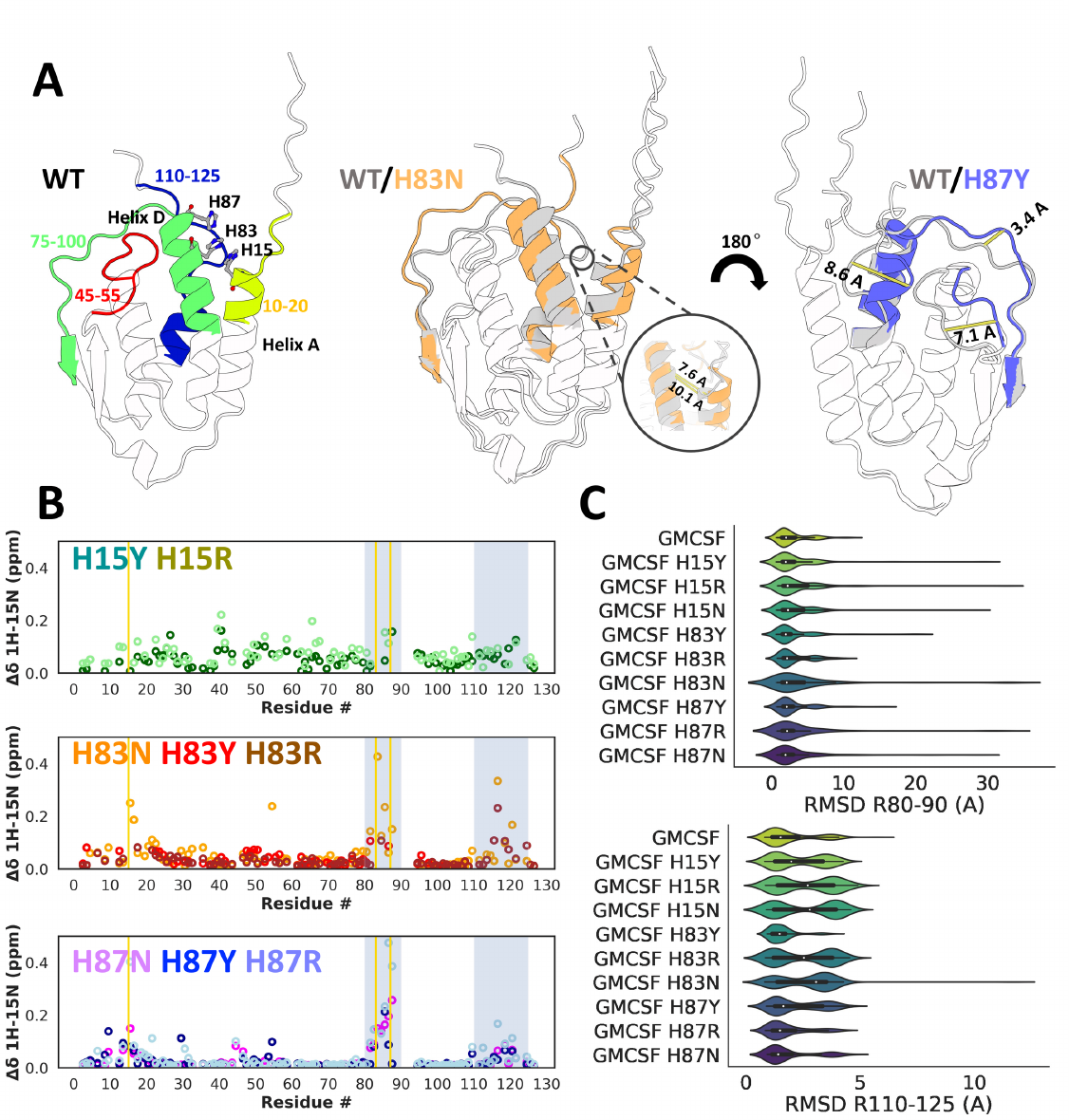}
    \caption{Subsampled AF2 results for GMCSF mutations. (A) Left: Annotated wild-type GMCSF structure in the ground (closed) conformation as predicted by AF2. Each colored element represents a region of putative high mobility according to chemical shift perturbations (CSPs) in the histidine triad mutants. Middle: Superposition of GMCSF in the ground conformation and a putative open conformation whose population is enriched in the AF2 predictions for H83 and H87 mutants, especially H83N. The inset compares distances between codon positions 15 and 83 in both conformations. Right: Superposition of GMCSF in the ground conformation and a tilted helix D conformation whose population is enriched in the AF2 predictions for H87 mutants (especially H87Y). Distances between flexible elements for both conformations are shown as yellow bars. (B) Per-residue chemical shift perturbations for GMCSF mutants, separated by codon position. Gold vertical bars represent the three sites of mutation, and silver shaded areas correspond to the residue indices plotted on the x-axis in C. (C) Distributions of root mean square deviations of atomic positions for the backbone atoms of residues 83-100 (top) and 110-125 (bottom) superimposed upon the GMCSF wild-type ground state structure. The distributions span 480 independent predictions (32 unique seeds * 5 models * 3 replicates.)}
     \label{fig_9}
\end{figure}

Further NMR experiments (Figure S6) have shown that point mutations in the aforementioned histidine triad lead to significant chemical shift perturbations, similar to those caused by reductions in pH \cite{Cui_2020}, hinting at increased occupancy of the open state or other unknown conformations. Importantly, the amplitudes of the changes to the backbone dynamics caused by these mutations vary significantly depending on which histidine of the triad is mutated. Specifically, mutating H15 or H83 leads to more pronounced changes in chemical shifts than mutating H87 (see Figure \ref{fig_9}). This is expected as H15 and H83 are significantly more buried than H87, meaning that mutations at these positions are likely to be harder for the GMCSF backbone to accommodate. Conversely, changes in the mostly solvent-exposed H87 usually lead to modest shifts in backbone dynamics, suggesting that mutations in this codon are more readily tolerated. Finally, mutations H83Y and H83R lead to larger scale conformational changes (inferred from broadened peaks) than any other mutation in the test set, and considerably more than the H83N mutation, suggesting that specific substitutions at each position induce more significant changes in the GMCSF backbone dynamics (see Figure S6 for a detailed analysis of all mutant CSPs and broadened peaks). These mutations are therefore useful for benchmarking as they represent a tiered challenge of predicting which specific amino acid substitution will lead to the largest changes in GMCSF structure.

Accordingly, we sought to test if our subsampled AF2 method was capable of predicting the expected changes in the backbone dynamics of each mutant with respect to the flexibility of helix A and other structural elements in GMCSF. Our predictions were considered accurate if they matched NMR experiments we performed along two axes: if they predicted mutations to H15 and H83 to provoke larger changes in the distribution of conformations than mutations to H87; and if they correctly predicted the most significant mutations in H83 or H87 as evidenced by
CSPs or broadened peaks. After building the master MSA using the wild-type human GMCSF sequence as a query and the JackHMMR method \cite{Finn_2011}, we determined the \emph{\emph{max\_seq}} and \emph{extra\_seq} parameters that led to the greatest diversity of GMCSF conformations while still maintaining the ground state as the most frequent state. As shown in Figure S7, \emph{max\_seq}=4 and \emph{extra\_seq}=8 were found to maximize diversity. Employing these parameters while keeping all others the same from the Abl1 tests, we used AF2 to predict wild-type GMCSF structures. We then repeated every previous step minus the parameter optimization for each GMCSF mutant and quantified the ground state populations from the resulting set of predictions.

To assess how mutations affected the relative populations of different GMCSF conformations, we measured the RMSD of specific backbone atomic positions of each predicted GMCSF structure with respect to the AF2 prediction that was most similar to the wild-type crystallographic reference (PDB 1CSG) \cite{Walter_1992}. The regions mentioned above are described in Figure \ref{fig_9} and correspond to GMCSF elements known to show significant perturbations in NMR experiments. 
The results of this analysis for the two regions with the most significant changes in our NMR results are described in Figure \ref{fig_9}. Additionally, we also measured the distance between the alpha carbons of H15 and H83 as well as the overall backbone RMSD with respect to the ground state reference for each prediction with the goal of identifying predictions that led to unexpected conformations or partially or completely unfolded structures. These additional measurements as well as three examples of unexpected structures are illustrated in Figure S8.

Despite the paucity of sequence data, our subsampled AF2 method correctly identified residues H15 and H83 as the most sensitive to mutation, as the range of the distribution of RMSDs of residues 80-90 and 110-125 is significantly larger for most of the mutations tested at both of these sites. Conversely and in-line with the experimental results, mutations to H87 led to significant changes in the distribution of the residue 80-90 RMSDs, but to comparatively modest changes for the RMSDs of residues 110-125. In addition to accurately predicting the differences in amplitude of backbone dynamics between mutants H15/83 and H87, subsampled AF2 correctly estimated the significant impact of mutations H83R and H83N on C-terminal dynamics, while also accurately predicting H83N, H83Y, and all three H87 mutants to have a large impact on the R80-90 RMSD distribution.

As with the Abl1 allelic series example, our method did not achieve perfect accuracy. Specifically, it failed to predict the significant changes in GMCSF's N-terminus dynamics associated with the H83N mutation (Figure S9). Additionally, our method failed at replicating the significant changes in dynamics of residues 80-90 for the H83Y mutations, which should be of larger amplitude than those for H83N and closer to those for H83R. Similarly, we expect the H87Y mutation to induce a larger amplitude of conformational changes in the region composed of residues 80-90 than the H87N mutation due to the enrichment of broadened peaks in that area in the H87Y NMR results, but AF2 predicts H87Y to be significantly less impactful than H87N in that context. In summary, AF2 performed exceptionally well at predicting subtle conformational changes in specific loops for most of our test cases but failed at replicating the comparatively larger expected effects of mutation H83Y.

Beyond these limitations, we also observed a set of novel conformations that are significantly different than both the ground and open states. Clustering the results by the structural features described in Figure S9 reveals that one alternative conformation in particular is significantly enriched, especially in predictions of the H83 mutants. In this alternative conformation, henceforth dubbed A1, the C helix has switched places with the B helix, placing H83 and H15 more than 10 \r{A} away (Figure S8). In this state, helix B occupies the groove to which heparin is thought to bind, which could be a mechanism for self-inhibition. Although further NMR experiments and simulations are necessary to confirm if this in fact a metastable GMCSF conformation with physiological functions, the fact that it has structural rearrangements with amplitudes that seem to be comparable to those of the H83 mutations (as evidenced by the H3 NMR chemical shift perturbations and broadened peaks [Figure S6]) and that this conformation was predicted more often for H83 mutants highlights the promise of using subsampled AF2 to help understand NMR results and derive novel hypotheses or mechanisms.

Finally, many of the alternative conformations not discussed above were found to be partially unfolded. Although the frequency of these was extremely low compared to predictions binned to the ground, open, or A1 states, the fact that they existed at all was surprising as running subsampled AF2 for the Abl1 test case led to no unfolded predictions whatsoever. One plausible hypothesis beyond the destabilizing effects of the mutations that may explain this observation is that the substantially shorter length of the GMCSF MSA relative to Abl1's drastically increases the uncertainty in AF2's predictions leading to a wider range of predictions.

\section{Discussion \label{conclusions}}

In this work, we successfully modified AF2's inputs and parameters to predict the conformational ensembles and relative state populations of two proteins, Abl1 kinase and human granulocyte-macrophase colony-stimulating factor (GMCSF), that have very different amounts of known sequence homology. In studying these proteins, we focused on how well our subsampled version of AF2 can reproduce the effects of evolution and mutations on relative state populations. 

Our subsampled AF2 implementation predicted Abl1, a kinase for which there is an abundance of sequence homology, to occupy its ground state nine times more frequently than its I2 state, consistent with NMR observations. More importantly, subsampled AF2 correctly predicted the relative state populations of two evolutionarily-related kinases, Anc-AS and Src kinase, leading to the correct correlation between their evolutionary distances and relative state populations. Using single and double Abl1 mutants with known effects on the relative state populations of Abl1's ground and I2 states, we moreover found that AF2 yielded surprisingly accurate state populations even for single mutations. This is best evidenced in the results for the F382 substitutions: phenylalanine 382 is a codon known to slightly increase the population of Abl1's ground state if mutated to leucine, or significantly reduce it if mutated to valine, an observation that is accurately replicated by our prediction method. Furthermore, an unexpected but remarkable feature illustrated by all of the Abl1 test cases is the capacity of subsampled AF to predict intermediate conformations spanning the transition from the ground state to I2 which closely match intermediate conformations obtained from more costly MD simulations.

Finally, we also obtained overwhelmingly accurate backbone dynamics predictions for GMCSF. Despite the lack of GMCSF sequence data which led to an input MSA of fewer than 120 sequences (versus more than 600,000 for Abl1), predictions from subsampled AF2 were distributed with frequencies that matched the expected relative state populations for most GMCSF variants. These results suggest that AF's prediction engine is robust enough to decode population signals from relatively scarce data.

The results we have obtained for the two test cases discussed above confirm AF2's potential for predicting conformational ensembles and, more importantly, demonstrate entirely novel applications of AF. In particular, we show that optimization of subsampling parameters allows users to accurately predict relative state populations and how they change in response to mutations with single-substitution resolution. This feature is a significant advance over the previous state of the art, as it facilitates high-throughput applications such as the design of fold-switching proteins, inference of mechanisms of acquired drug resistance, and reweighting of binding affinity predictions. Additionally, our workflow generates conformational intermediates, which has direct implications for discovering  drugs that bind alternative conformations and for improving our understanding of biophysics in general. In addition to these immediate practical applications, the more fundamental observation that AF2 is decoding information regarding conformational distributions from sequence data alone points to many other potential unforeseen uses of AF2 that can result in further methodological advances and discoveries. 

This said, our AF2 pipeline is not perfect; our workflow inaccurately predicted the M290L mutation to significantly decrease the ground state population when that mutation is known experimentally to have the opposite effect. Interestingly, AF2 predicted the double mutation (M290L+L301I) to increase the ground state population more than the L301I mutation alone. As AF2 correctly predicted the relative state populations of Src, which differs from Abl1 by dozens of mutations, this suggests a potential more general trend that AF2 is more accurate in predicting the effect of multiple mutations than that of a single mutation. Furthermore, our pipeline also struggled to correctly predict all of the structural elements expected to differ in each conformation. Specifically, while AF2 predicted an ensemble of activation loop conformations, a few of the inactive-like predictions (activation loop closed) contained structural elements that are typically thought to belong to the active state, such as the position of the C-helix and the dihedral angles of the D381 and F382 side-chains.
Moreover, even when our pipeline predicts a change in a structural element from its configuration in the ground (active) state, the amplitude of that change may not correspond to that seen experimentally. For example, the side-chain dihedral angle of the D381 residue in Abl1 ranges from -130 (active state) to 40 (inactive state) degrees in structures in the Protein Data Bank (PDB) \cite{Xie_2020}, but ranges from -130 to -90 degrees in AF2 predictions. Last but not least, while our method is capable of predicting whether mutations will increase or decrease state populations, it remains to be shown whether it is capable of directly predicting Boltzmann ratios of states in a quantitative fashion. 

Despite these limitations, we believe that our results showcase promising, unexpected applications of AF2. Using the high-throughput pipeline we developed or one inspired by it could save significant time when filtering large allelic series to identify mutations with significant impacts on conformation. It could also be used as a prediction engine for classifying arrays of drug-resistant mutations based on their shared effects on the stability of a given conformation, thus facilitating polypharmacology. Finally, as demonstrated by AF2's Abl1 activation loop predictions, our method could also be useful for identifying previously unknown, potentially metastable states of known proteins. While it remains to understand exactly how AF2 is gathering and interpreting signals about state populations from sequence data, we hope that our work will motivate many future investigations. 

\section{Methods \label{Methods}}

\subsection{Protein Structure Visualization \label{visualization}}
To visualize the predicted structures and trajectories and calculate descriptors such as distances between atoms, RMSD to reference, dihedral angles, etc., we used PyMol (version 2.4.1) (Schrodinger LLC, 2020).

\subsection{Multiple Sequence Alignments \label{MSAs}}
The JackHMMR algorithm \cite{Finn_2011} was used to generate multiple sequence alignments (MSAs) for each protein of interest by querying sequences from the UniProt90 \cite{Suzek_2014}, Mgnify \cite{Richardson_2022}, and BFD \cite{Steinegger_2018} databases.

\subsection{Structure Prediction \label{prediction}}
We used AlphaFold 2 \cite{jumper_highly_2021} within the localcolabfold colabfold-batch 1.5.0 implementation  \cite{Mirdita2022} to predict protein structures of Abl1, Src, ANC-AS, and of GMCSF.

\subsection{Molecular Dynamics Simulations \label{md}}
Enhanced-sampling molecular dynamics simulations were conducted on the proteins described using the WESTPA2 \cite{Bogetti_2022} implementation within the OpenMM molecular simulation engine \cite{Eastman_2017}. For an extensive description of the MD methodology employed in this study, please refer to the Supplementary Materials.

\subsection{Protein Expression and Purification \label{nmr1}}

Plasmid DNA containing either wildtype GMCSF or GMCSF containing a point mutation with an N terminal 6-His tag was cloned into a pET-15b vector and transformed into BL21(DE3) cells in a manner described elsewhere. Isotopically enriched GMCSF was expressed at 37\textdegree C in M9 minimal medium containing CaCl$_2$, MgSO$_4$ and MEM vitamins with $^1$$^5$NH$_4$Cl as the sole nitrogen source. Small cultures of GMCSF were grown overnight in LB medium. The following morning, cloudy suspensions were collected by centrifugation and resuspended in the final M9 growth medium. Cultures of GMCSF were grown to an OD$_6$$_0$$_0$ of 0.8-1.0 before induction with 1 mM isopropyl $\beta$-D-1- thiogalactopyranoside (IPTG). Post induction, cells were kept at 18\textdegree C while shaking. Cells were harvested after 18h and resuspended in a denaturing lysis buffer containing 10 mM Tris-HCl, 100 mM sodium phosphate, and 6 M guanidine hydrochloride (GuHCl) at pH 8.0. Cells were lysed by sonication and cell debris was removed by centrifugation. The resulting supernatant was incubated and nutated with 10 mL of Ni-NTA agarose beads for 30 min at room temperature before the Ni-NTA slurry was packed into a gravity column. The column was washed with the initial lysis buffer, followed by a gradient of the same buffer without GuHCl over 100 mL. Elution of GMCSF in its denatured form was performed with 1 column volume of a buffer containing 10 mM Tris-
HCl, 100 mM sodium phosphate, and 250mM imidazole at pH 8.0.
GMCSF was refolded by dilution via dropwise addition of the 10 mL elution into 100 mL of a refolding buffer containing 10mM Tris-HCl, 100mM sodium phosphate, and 750 mM arginine at pH 8.0. The refolded protein was dialyzed exhaustively against a buffer containing 2 mM sodium phosphate at pH 7.4. GMCSF was concentrated to $\sim$200 $\mu$M with an Amicon centrifugal device and stored at -20\textdegree C.

\subsection{NMR Spectroscopy \label{nmr2}}
NMR samples were prepared by dialyzing 200 $\mu$ GMCSF and GMCSF mutants
against a buffer of 20 mM HEPES and 1 mM EDTA at pH 7.4. NMR experiments were performed on a Bruker Avance NEO 600 MHz spectrometer at 25\textdegree C. NMR data were processed in NMRPipe \cite{Delaglio1995} and analyzed in Sparky \cite{Lee2014}. The $^{1}$H and $^{15}$N carrier frequencies were set to water resonance and 120 ppm, respectively.

\subsection{Miscellaneous and Data Visualization \label{misc}}
Data plotting and visualization were performed using Seaborn (version 0.11.1). Data were plotted as mean ± confidence intervals.

\section{Acknowledgements \label{ack}}
The authors thank Jacob Rosenstein, F. Marty Ytreberg, Jagdish S. Patel, Marcelo D. Polêto, Winston Y. Li, Haibo Li, and Kyle Lam for insightful conversations and support. G.M.d.S. and B.M.R. were supported in part by the National Science Foundation under Grant No. 2027108. J.C. and G.L.'s contributions were supported by NIH Grant R01 GM144451. The computational aspects of this research were conducted using computational resources and services at the Center for Computation and Visualization, Brown University.

\section{Author Contributions}
G.M.d.S. performed the simulations and analyzed the sequence and AlphaFold 2 data. J.C. performed NMR experiments and contributed the NMR data. 
D.D., G.L., and B.M.R. provided direction and oversight.
G.M.d.S. and B.M.R. drafted the manuscript. All authors edited the manuscript.

\section{Competing Interests}
There are no competing interests. 

\section{Data Availability}
The datasets from this study are available from the authors upon request.

\section{References}
\printbibliography[heading=none]

\end{document}


\include{MyCommand}
\maketitle

\maketitle

\section{Abl1 Ortholog Sequences Used to Generate Multiple Sequence Alignments}

\begin{figure}[H]
\centering
    \includegraphics[width=450pt]{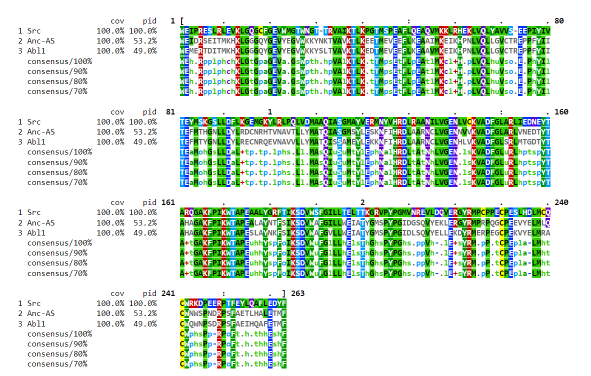}[width=450pt]
    \caption{Sequences of the Abl1, Src, and Anc-AS kinase cores used to generate MSAs and as
input for subsampled AlphaFold 2.}
     \label{Fig S1}
\end{figure}

\section{Comparisons Between AF2 Predictions and Molecular Dynamics Snapshots along the Ground to I2 Transition}

\begin{figure}[H]
\centering
    \includegraphics[width=450pt]{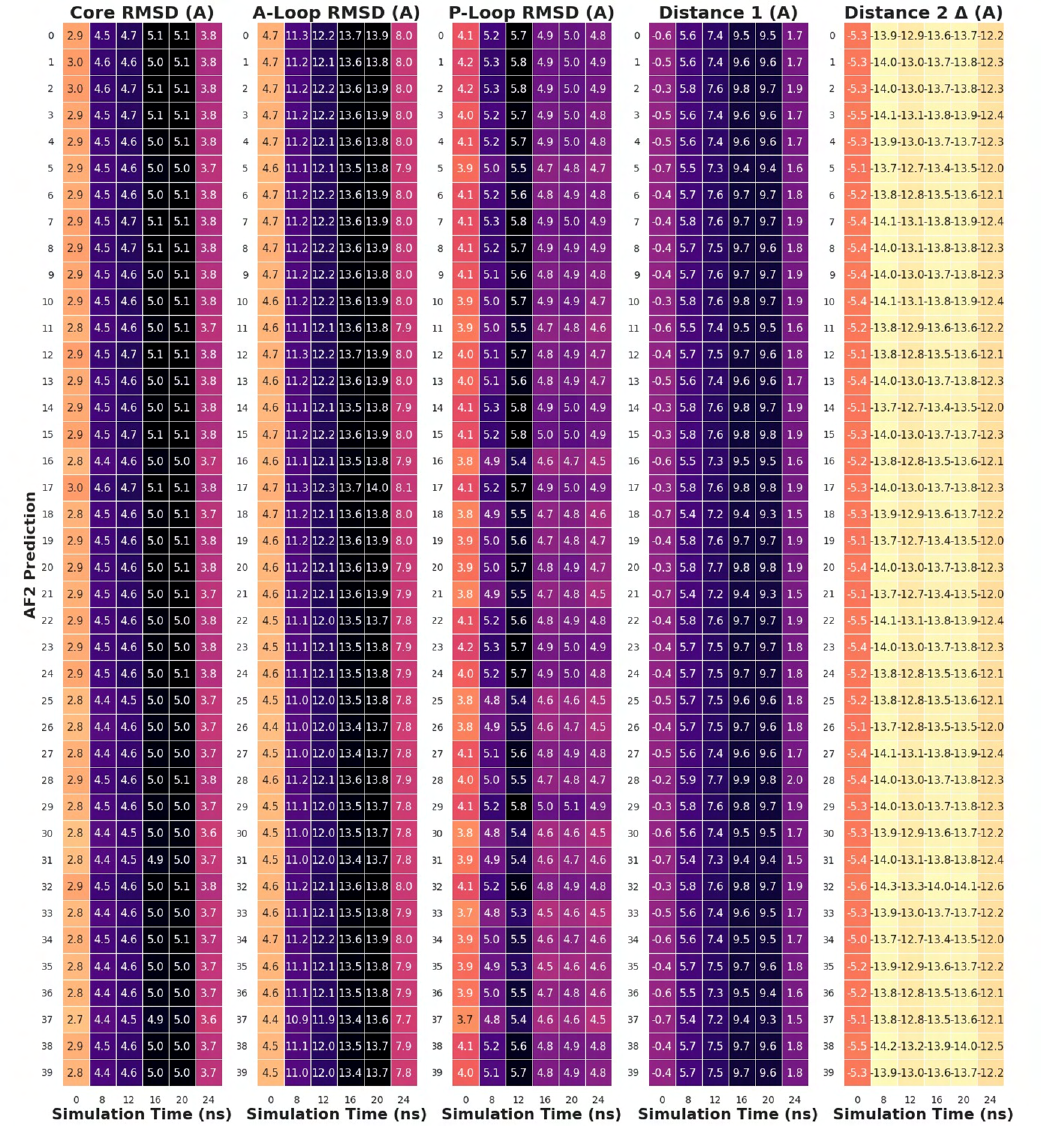}
    \caption{Part one of four of the comparison between the values of five structural elements in the Abl1 kinase core known to change during the ground to I2 transition as measured from the ensemble of 160 subsampled AF2 predictions and six frames extracted from a molecular dynamics simulation trajectory spanning the transition at different time points. Core, P-Loop, and A-Loop RMSDs are defined as the backbone RMSDs of each AF2 prediction's kinase core (residues 242 to 459), activation loop (residues 379 to 395), or phosphate-binding loop (residues 244 to 256)  vs. the kinase core, phosphate-binding loop, or activation loop backbone of the MD snapshot selected at each time point. Distance deltas are defined as the difference in atom pair distances between each AF2 prediction and its respective MD snapshot. Distance 1 corresponds to the distance between the backbone oxygens of E377 and L409, and Distance 2 corresponds to the distance between the backbone oxygens of L409 and G457.}
     \label{Fig S2}
\end{figure}

\begin{figure}[H]
\centering
    \includegraphics[width=450pt]{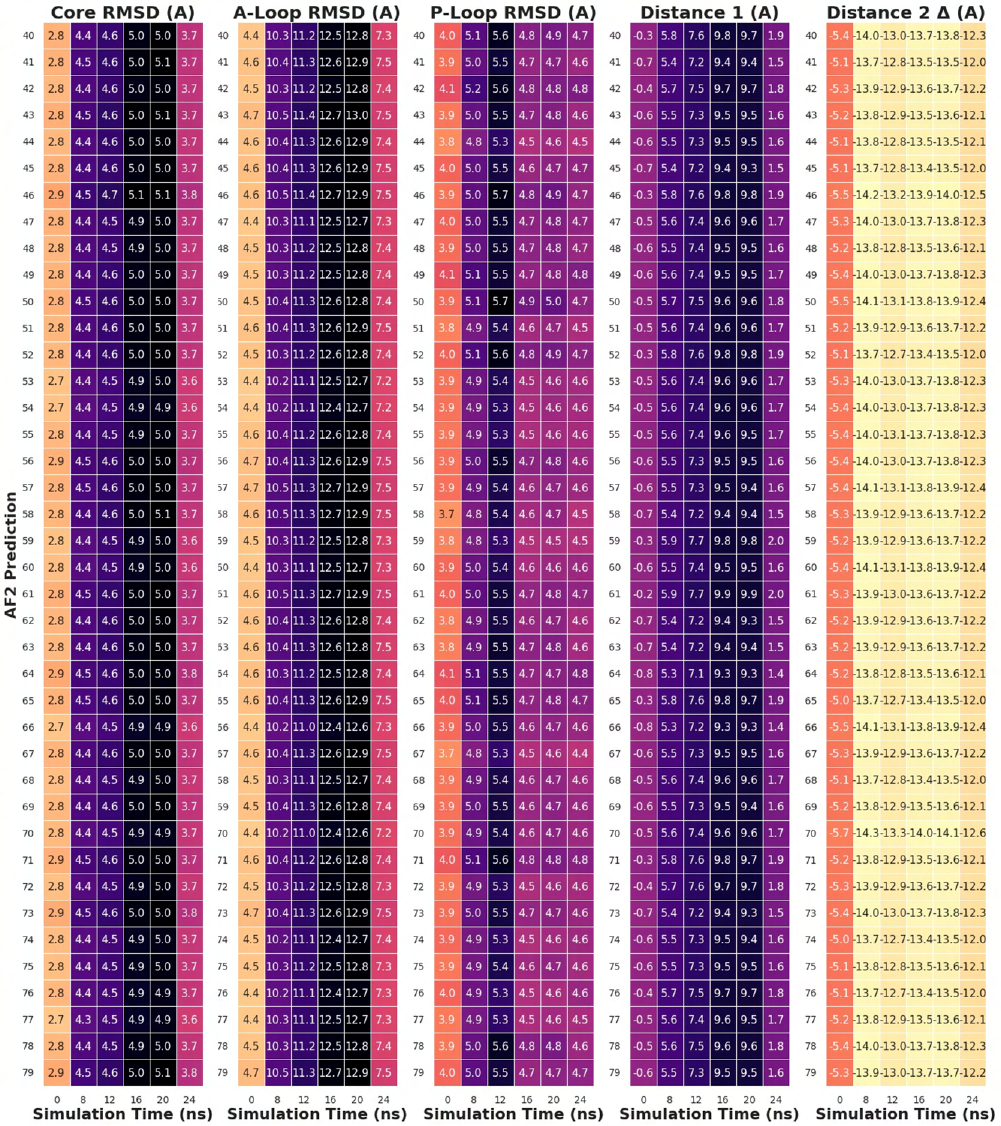}
    \caption{Part two of four of the comparison between the values of five structural elements in the Abl1 kinase core known to change during the ground to I2 transition as measured from the ensemble of 160 subsampled AF2 predictions and six frames extracted from a molecular dynamics simulation trajectory spanning the transition at different time points. Core, P-Loop, and A-Loop RMSDs are defined as the backbone RMSDs of each AF2 prediction's kinase core (residues 242 to 459), activation loop (residues 379 to 395), or phosphate-binding loop (residues 244 to 256)  vs. the kinase core, phosphate-binding loop, or activation loop backbone of the MD snapshot selected at each time point. Distance deltas are defined as the difference in atom pair distances between each AF2 prediction and its respective MD snapshot. Distance 1 corresponds to the distance between the backbone oxygens of E377 and L409, and Distance 2 corresponds to the distance between the backbone oxygens of L409 and G457.}
     \label{Fig S3}
\end{figure}

\begin{figure}[H]
\centering
    \includegraphics[width=450pt]{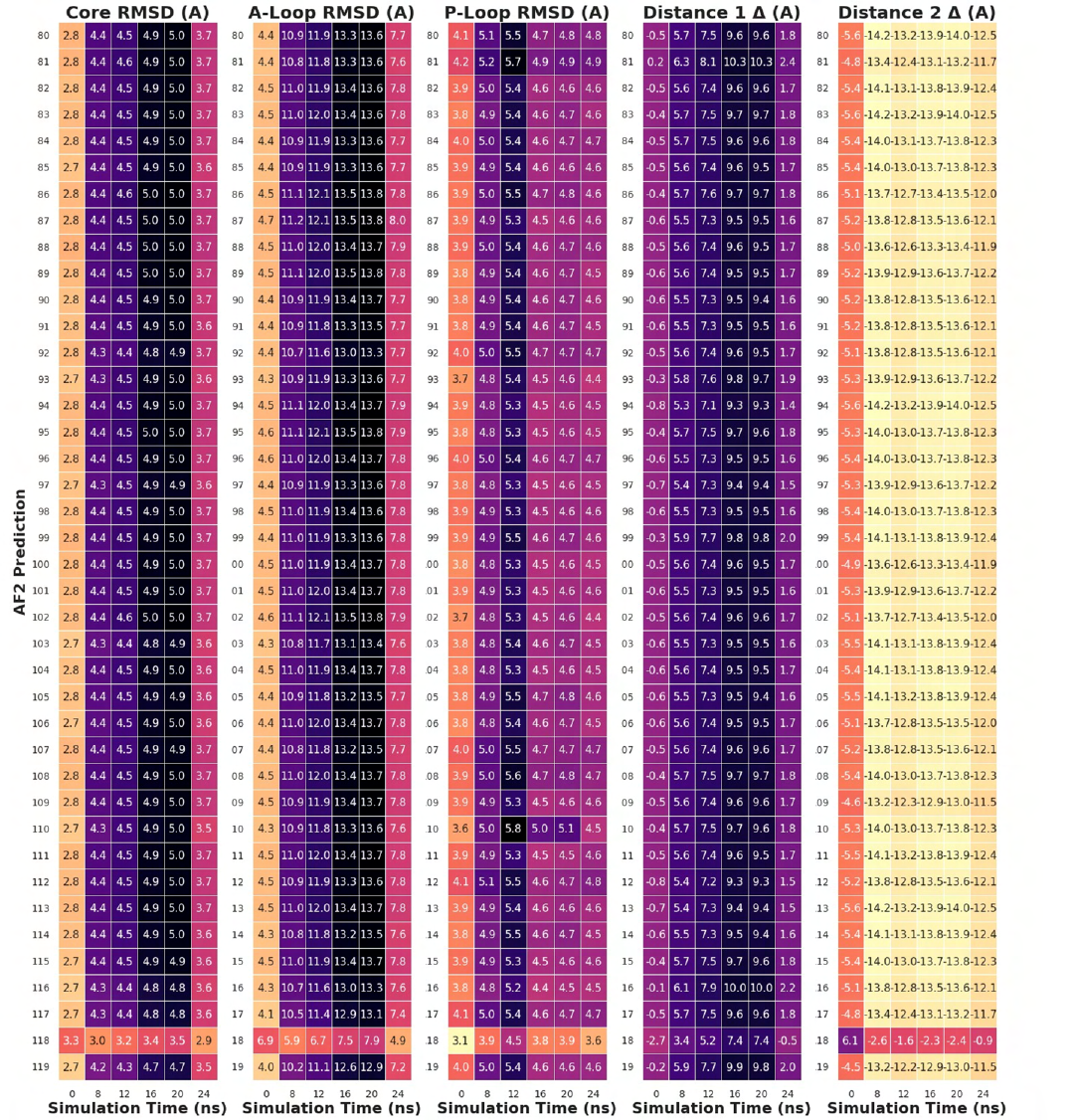}
    \caption{Part three of four of the comparison between the values of five structural elements in the Abl1 kinase core known to change during the ground to I2 transition as measured from the ensemble of 160 subsampled AF2 predictions and six frames extracted from a molecular dynamics simulation trajectory spanning the transition at different time points. Core, P-Loop, and A-Loop RMSDs are defined as the backbone RMSDs of each AF2 prediction's kinase core (residues 242 to 459), activation loop (residues 379 to 395), or phosphate-binding loop (residues 244 to 256)  vs. the kinase core, phosphate-binding loop, or activation loop backbone of the MD snapshot selected at each time point. Distance deltas are defined as the difference in atom pair distances between each AF2 prediction and its respective MD snapshot. Distance 1 corresponds to the distance between the backbone oxygens of E377 and L409, and Distance 2 corresponds to the distance between the backbone oxygens of L409 and G457.}
     \label{Fig S4}
\end{figure}

\begin{figure}[H]
\centering
    \includegraphics[width=450pt]{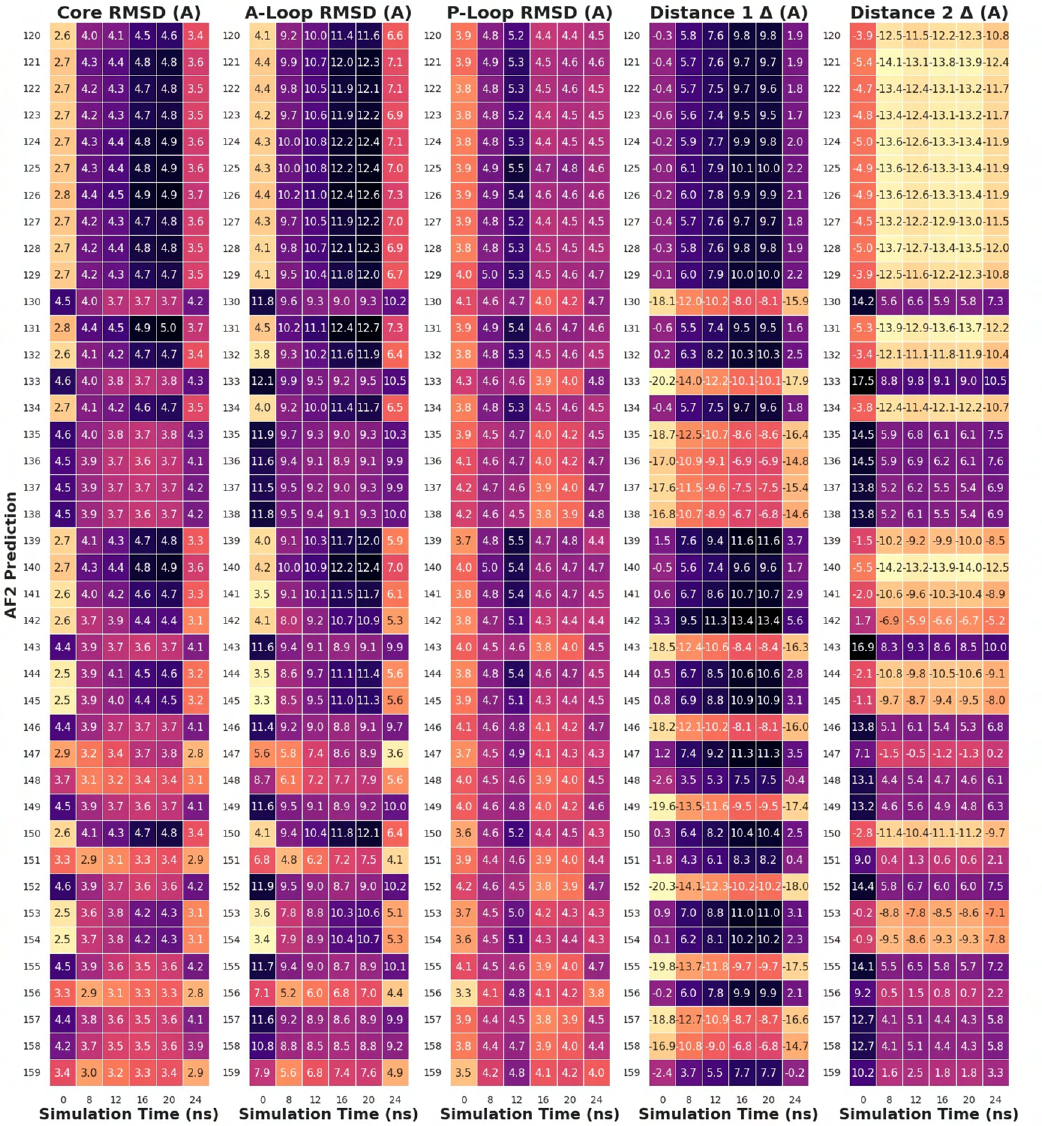}
    \caption{Part four of four of the comparison between the values of five structural elements in the Abl1 kinase core known to change during the ground to I2 transition as measured from the ensemble of 160 subsampled AF2 predictions and six frames extracted from a molecular dynamics simulation trajectory spanning the transition at different time points. Core, P-Loop, and A-Loop RMSDs are defined as the backbone RMSDs of each AF2 prediction's kinase core (residues 242 to 459), activation loop (residues 379 to 395), or phosphate-binding loop (residues 244 to 256)  vs. the kinase core, phosphate-binding loop, or activation loop backbone of the MD snapshot selected at each time point. Distance deltas are defined as the difference in atom pair distances between each AF2 prediction and its respective MD snapshot. Distance 1 corresponds to the distance between the backbone oxygens of E377 and L409, and Distance 2 corresponds to the distance between the backbone oxygens of L409 and G457.}
     \label{Fig S5}
\end{figure}

\section{GMCSF Chemical Shift Perturbations}
\begin{figure}[H]
\centering
    \includegraphics[width=450pt]{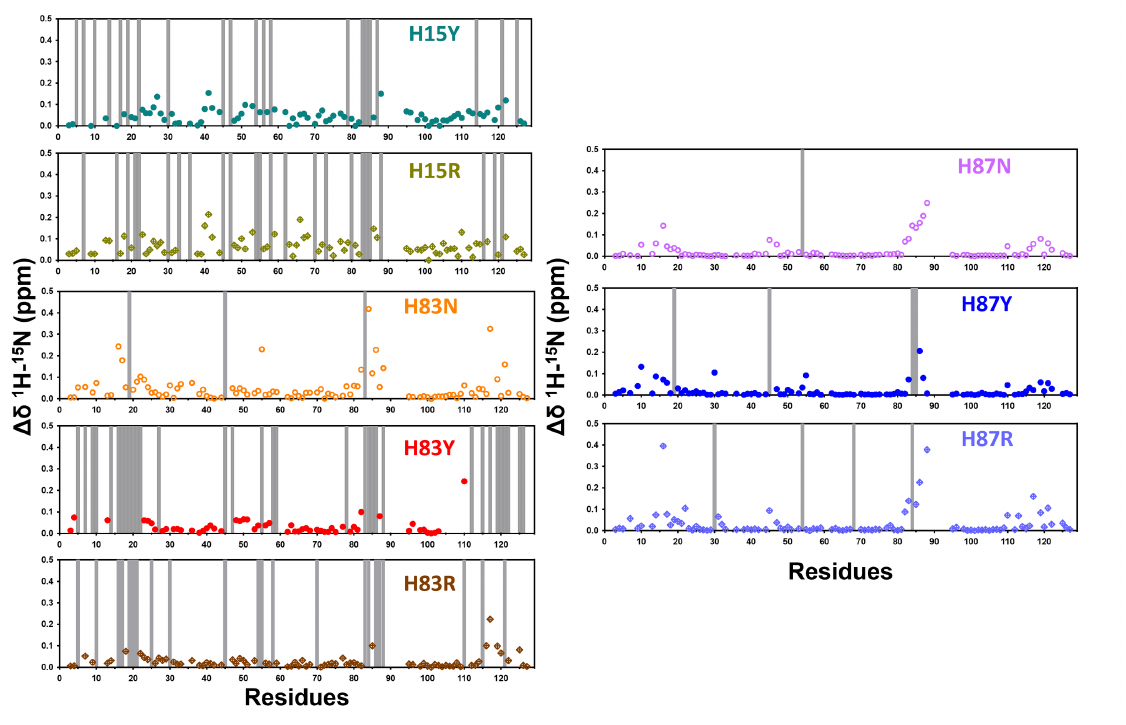}
    \caption{N15-H1 Chemical shift perturbations for mutant GMCSF constructs in reference to wild-type GMCSF peaks. Vertical bars indicate residues where the signal was lost due to chemical exchange broadening.}
     \label{Fig S6}
\end{figure}

\section{Optimization of AF2 Parameters for the GMCSF Protein}

\begin{figure}[H]
\centering
    \includegraphics[width=450pt]{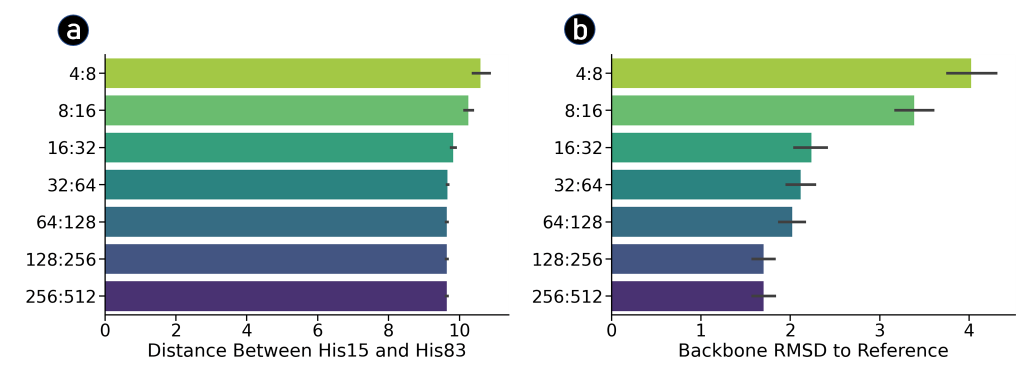}
    \caption{Optimal AF2 subsampling parameters for GMCSF. (a) Effects of modifying the \textit{max\_seqs} and \textit{extra\_seqs} values on the diversity of the distances between the H15 and H83 residues observed, which is a proxy for the opening of the heparin-binding site in GMCSF. (b) Effects of modifying the \textit{max\_seqs} and \textit{extra\_seqs} values on the diversity of the root mean square deviation of atomic positions (RMSD) of the GMCSF backbone with respect to the ground state reference (PDB 1CSG).}
     \label{Fig S7}
\end{figure}

\section{GMCSF Dynamics Prediction}

\begin{figure}[H]
\centering
    \includegraphics[width=450pt]{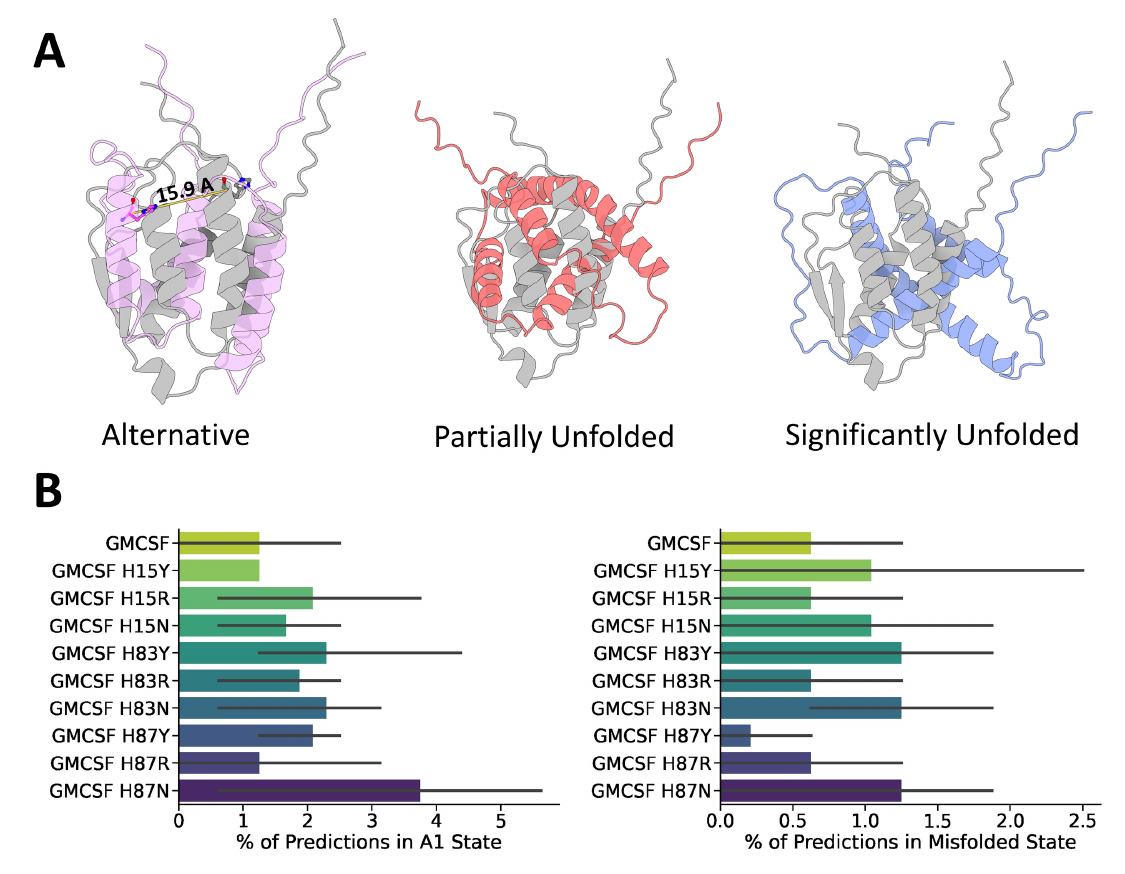}
    \caption{Unusual GMCSF states predicted by subsampled AF2 and the respective populations of those states. (a) Structure of the most common alternative state predicted by AF2 (A1, in pink) aligned with and overlain on a ground state prediction (in grey). The distance between H83 in the reference and in conformation A1 is displayed as a measure of the difference between the conformations. Also shown are two misfolded/unfolded predictions aligned with and overlain on the ground state prediction (in grey). (b) AF2 predictions of the relative populations of the A1 conformation and the misfolded/unfolded structures. Conformations were classified as the A1 conformation based on the distance between the H15 and H83 residues (greater than 11 \r{A}) and overall backbone RMSD vs. ground state reference (greater than 5 A but less than 10 \r{A}), while they were classifies as misfolded/unfolded based on overall backbone RMSD vs. the ground state reference (equal to or greater than 13 \r{A}).}
     \label{Fig S8}
\end{figure}

\begin{figure}[H]
\centering
    \includegraphics[width=450pt]{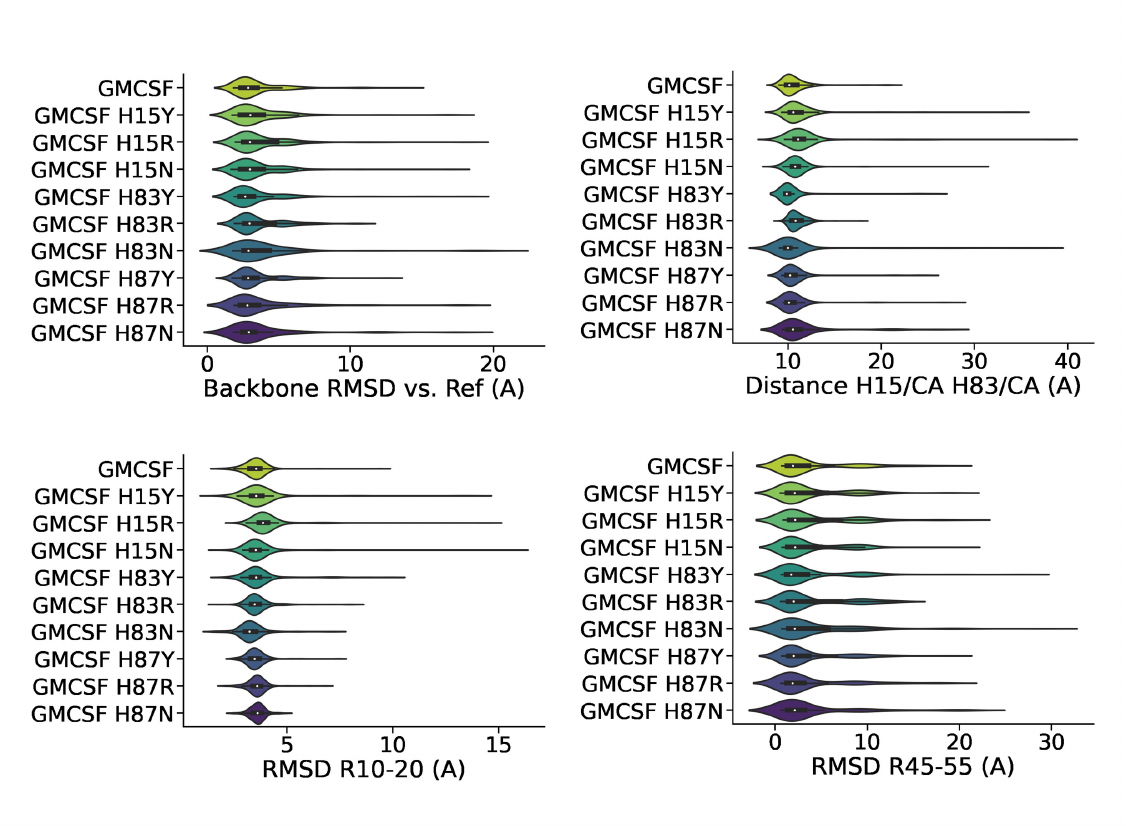}
    \caption{AF2 predictions of the distributions of different GMCSF properties. Every RMSD measurement was taken with respect to the ground state reference (PDB 1CSG).}
     \label{Fig S9}
\end{figure}

\section{Optimization of AF2 Parameters for the Abl1 Protein}

\begin{table}[H]
    \centering
    \caption{Optimized AF2 parameters for predicting Abl1 ensembles.} 
    \begin{tabular}{lllllll}
        \textbf{parameter\_test} & \textbf{max\_seq} & \textbf{extra\_seq} & \textbf{n\_recycles} & \textbf{n\_models} & \textbf{n\_seeds} & \textbf{\%\_notground} \\ \hline
        \textbf{t\_max\_extra\_1} & 32 & 64 & 4 & 5 & 32 & 2 \\ 
        \textbf{t\_max\_extra\_2} & 64 & 128 & 4 & 5 & 32 & 5 \\ 
        \textbf{t\_max\_extra\_3} & 128 & 256 & 4 & 5 & 32 & 9 \\ 
        \textbf{t\_max\_extra\_4} & 256 & 512 & 4 & 5 & 32 & 18 \\ 
        \textbf{t\_max\_extra\_5} & 512 & 1024 & 4 & 5 & 32 & 15 \\ 
        \textbf{t\_max\_extra\_6} & 2048 & 4096 & 4 & 5 & 32 & 7 \\ 
        \textbf{t\_max\_extra\_7} & 4098 & 8192 & 4 & 5 & 32 & 6 \\ 
        \textbf{t\_max\_extra\_8} & 512 & 32 & 4 & 5 & 32 & 18 \\ 
        \textbf{t\_max\_extra\_9} & 32 & 512 & 4 & 5 & 32 & 1 \\ 
        \textbf{t\_nseeds\_1} & 256 & 512 & 4 & 5 & 128 & 12 \\ 
        \textbf{t\_nseeds\_2} & 256 & 512 & 4 & 5 & 300 & 12 \\ 
        \textbf{t\_nrecycles\_1} & 32 & 64 & 8 & 5 & 128 & 0 \\ 
        \textbf{t\_nrecycles\_2} & 32 & 64 & 8 (kept) & 5 & 128 & 2 \\ 
        \textbf{t\_nrecycles\_3} & 256 & 512 & 8 & 5 & 128 & 8 \\ 
        \textbf{t\_nrecycles\_4} & 256 & 512 & 8 (kept) & 5 & 128 & 21 \\ 
    \label{Table S1}
    \end{tabular}
\end{table}

\section{AF2 Predictions of the Relative State Populations of Abl1 Kinase Core Mutants}

\begin{table}[H]
    \centering
    \caption{Abl1 kinase core mutants and their observed or expected effects on the relative populations of the active (Ground), inactive 1 (I1), or inactive 2 (I2) states.}
    \label{table:S2}
    \begin{tabular}{lllllll}
        \textbf{} & \textbf{Ground} & \textbf{I1} & \textbf{I2} \\ \hline
        \textbf{Wild-Type} & 88 & 6 & 6 \\ 
        \textbf{} & ~ & ~ & ~ \\ 
        \textbf{M290L} & 55 & 10 & 35 \\ 
        \textbf{L301I} & 25 & 10 & 65 \\ 
        \textbf{M290L + L301I} & 8 & 10 & 82 \\ 
        \textbf{} & ~ & ~ & ~ \\ 
        \textbf{F382L} & 90 & 0 & 10 \\ 
        \textbf{F382Y} & 10 & 0 & 90 \\ 
        \textbf{F382V} & 5 & 0 & 95 \\ 
        \textbf{} & ~ & ~ & ~ \\ 
        \textbf{I2M} & 10 & 0 & 90 \\ 
        \textbf{E255V (I2M background)} & nr & nr & 45 \\ 
        \textbf{T315I (I2M background)} & 93 & 0 & 7 \\ 
        \textbf{E255V + T315i} & nr & nr & nr
    \end{tabular}
\end{table}

\section{Molecular Dynamics and WESTPA2 Simulations\label{misc}}
Molecular dynamics simulations of wild-type Abl1 were conducted using the OpenMM software package (Eastman 2017) with the amber99sb-ildn force field (Lindorff Larsen 2010) and the tip3p water model (Mark 2001) at 300 K and 1 atm. The lowest energy Abl1 structure from the PDB 6XR6 NMR ensemble was solvated within a dodecahedron box and charges were neutralized by replacing a number of solvent atoms with chloride and sodium ions. Following solvation, we minimized the energy of each system using a steepest-descent algorithm until the maximum force on any given atom was less than 1000 kJ/mol/min or until 50,000 minimization steps were conducted. We ran the simulations with a 1 fs time step during the equilibration phase and a 2 fs time step during the production phase. We equilibrated solvent atoms first for 1 ns in the NVT ensemble and then for 1 ns in the NPT ensemble with solute heavy atoms restrained using the LINCS algorithm with a spring constant of 1,000 kJ/mol/m$^2$ (Hess 1997). The production phase (in the NPT ensemble) followed the equilibration phase but without restraints.

We used the WESTPA2 (Bogetti 2022) enhanced-sampling method to access the timescales necessary to simulate the inactivation pathway of Abl1. This was done via two WESTPA2 simulations (ground to I1 and I1 to I2). As progress coordinates for the ground to I1 transition, we defined the distance between the backbone oxygen of V299 and the center of mass of the carboxyl group of D381 as PC1; and the angle formed by the center of mass of the carboxyl group of D381, the backbone oxygen of K379, and the center of mass of the aromatic ring of F382 as PC2. For the I1 to I2 transition, we defined the distance between the backbone oxygen of L409 and the backbone oxygen of E377 as PC1; and the distance between backbone oxygen of L409 and the backbone oxygen of G4598 as PC2. Representative illustrations of the progress coordinates used in this protocol are in Figure S10, and their distributions and start/end state definitions are described in Fig S11. We ran WESTPA2 for 300 iterations for each leg of the transition, with the number of walkers per iteration varying from 64 to 512 due to the adaptive binning scheme, and 100 ps per iteration, totaling over 9 us of aggregate simulation time for each leg of the transition.

\begin{figure}[H]
\centering
    \includegraphics[width=450pt]{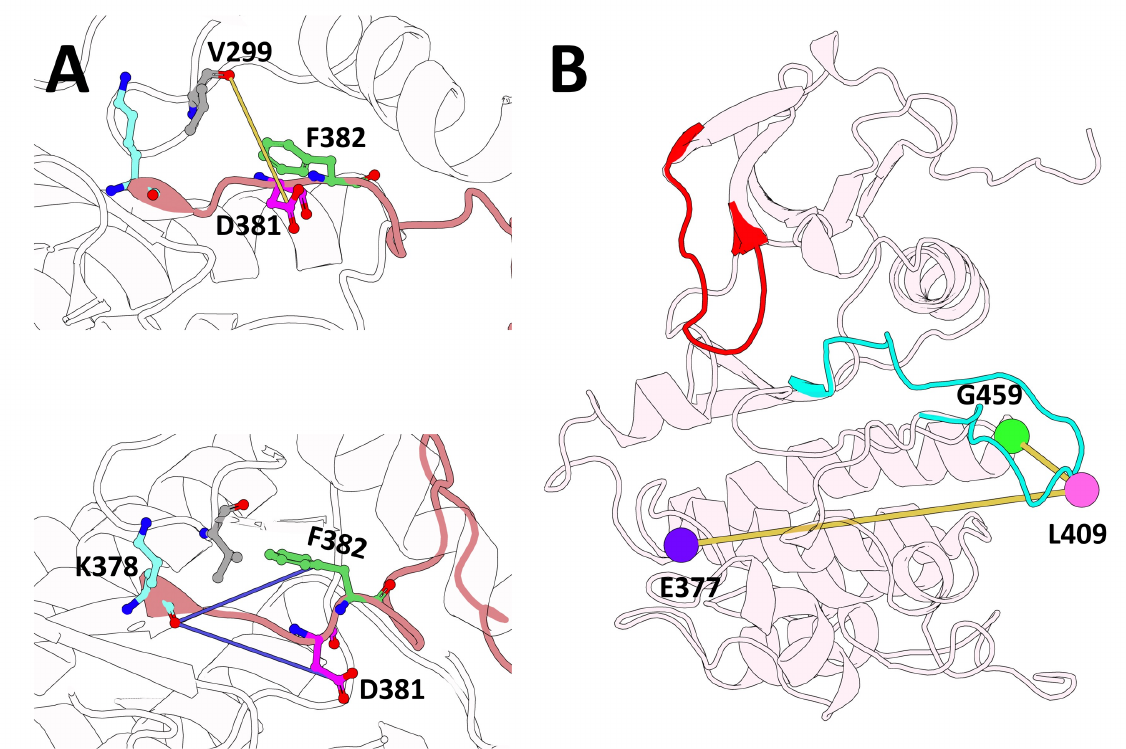}
    \caption{Progress coordinates used in the WESTPA2 simulations of wild-type Abl1. (A) Progress coordinates used in sampling the transition from the ground to the I1 state. (B) Progress coordinates used in sampling the transition from the ground to the I2 state. }
     \label{Fig 10}
\end{figure}

\begin{figure}[H]
\centering
    \includegraphics[width=450pt]{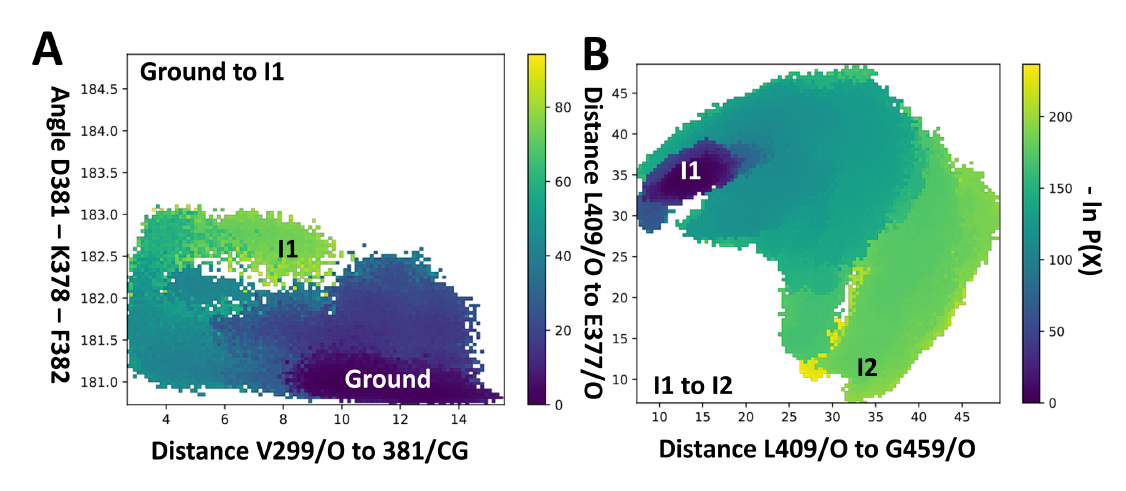}
    \caption{Distribution of values for the progress coordinates used in either the transition from the (A) ground to the I1 state or (B) I1 to I2 state.}
     \label{Fig S11}
\end{figure}